\documentclass[11pt]{article}

\usepackage{amssymb, latexsym, amsmath, amsthm}
\usepackage{amsfonts, graphics, amstext}
\usepackage{color}
\usepackage[demo]{graphicx}
\usepackage{caption}
\usepackage{subcaption}
\usepackage{float}

\def\R{\mathbb R}

\def\be{\begin{equation}}
\def\ee{\end{equation}}
\def\bea{\begin{eqnarray}}
\def\eea{\end{eqnarray}}
\def\beas{\begin{eqnarray*}}
\def\eeas{\end{eqnarray*}}

\newcommand{\prfe}{\hspace*{\fill} $\Box$
\smallskip \noindent}

\begin{document}

\sloppy

\newtheorem{theorem}{Theorem}[section]
\newtheorem{definition}[theorem]{Definition}
\newtheorem{proposition}[theorem]{Proposition}
\newtheorem{example}[theorem]{Example}
\newtheorem{cor}[theorem]{Corollary}
\newtheorem{lemma}[theorem]{Lemma}
\theoremstyle{remark}
\newtheorem*{remark}{Remark}

\renewcommand{\theequation}{\arabic{section}.\arabic{equation}}

\title{On the rotation curves for axially symmetric disk
       solutions of the Vlasov-Poisson system}

\author{H{\aa}kan Andr\'{e}asson\\
        Mathematical Sciences\\
        Chalmers University of Technology\\
        University of Gothenburg\\
        S-41296 G\"oteborg, Sweden\\
        email: hand@chalmers.se
       \\
        \ \\
        Gerhard Rein\\
        Fakult\"at f\"ur Mathematik, Physik und Informatik\\
        Universit\"at Bayreuth\\
        D-95440 Bayreuth, Germany\\
        email: gerhard.rein@uni-bayreuth.de}

\maketitle

\begin{abstract}
A large class of flat axially symmetric solutions to the Vlasov-Poisson system
is constructed with the property that the corresponding rotation curves are
 approximately flat, slightly decreasing or slightly increasing. The rotation
curves are compared with measurements from real galaxies and satisfactory
agreement is obtained. These facts raise the question whether the observed rotation
curves for disk galaxies may be explained without introducing dark matter.
Furthermore, it is shown that for the ansatz we consider
stars on circular orbits do not exist in the neighborhood of the boundary
of the steady state.

\end{abstract}

\section{Introduction}
\setcounter{equation}{0}

The rotation curve of a galaxy depicts the magnitude of the orbital velocities
of visible stars or gas particles in the galaxy versus their radial distance
from the center. In the pioneering observations by Bosma~(1981) and
Rubin et al~(1982) it was found that the rotation curves of spiral galaxies
are approximately flat except in the inner region where the rotation curves
rise steeply. Independent observations in more recent years agree with these
 conclusions. The flat shape of the rotation curves is an essential reason for
introducing the concept of dark matter. Let us cite from
(Famaey \& McGaugh 2012):
"Perhaps the most persuasive piece of evidence [for the need of dark matter]
was then provided, notably through the seminal works of Bosma and Rubin, by
establishing that the rotation curves of spiral galaxies are approximately
flat (Bosma~1981, Rubin et al~1982).
A system obeying Newton's law of gravity should have a
rotation curve that, like the Solar system, declines in a Keplerian manner
once the bulk of the mass is enclosed: $V_c\propto r^{-1/2}$."

The last statement is heuristic and it is therefore essential to construct
self-consistent mathematical models which describe disk galaxies and study the
corresponding rotation curves.
For this purpose it is natural to consider the Vlasov-Poisson system which
is often used to model galaxies and globular clusters. In fact, there exist
well-known explicit solutions to the Vlasov-Poisson system describing axially
symmetric disk galaxies which give rise to flat or even increasing rotation
curves. The Mestel disks and the Kalnajs disks are examples of such solutions,
cf.~(Binney \& Tremaine 1987).
However, these solutions are not considered physically
realistic. Mestel disks, which give rise to flat rotation curves, are infinite
in extent and their density is singular at the center. Schulz (2012) has
obtained finite versions of Mestel disks
but their density is still singular. In the case of the Kalnajs disks,
which give rise to linearly increasing rotation curves, there are convincing
arguments in (Kalnajs~1972)  that they are dynamically unstable.
This conclusion is supported by the numerical simulations we carry out as will
be discussed below.

The approach in the present investigation is different.
Our aim is first to construct solutions which are realistic
in the sense that they are dynamically stable,
have finite extent and finite mass, and then to study the corresponding
rotation curves.
However, the existence and stability theory for flat axially symmetric steady
states of the Vlasov-Poisson system is much less developed than
in the spherically symmetric case.
This limits our understanding of which flat steady states are realistic.
Nevertheless, motivated by the results in
(Rein~1999, Firt \& Rein~2006) we search for solutions
which we expect to be physically realistic, but we emphasize that the solutions
we construct are not covered by the present theory.

At this point a possible source of confusion concerning the definition
and interpretation of the rotation velocity for a given steady state of
the Vlasov-Poisson system must be addressed.
We define this velocity at radius $r$ as the velocity of a test particle
on a circular orbit of that radius in the gravitational potential of the
steady state, provided such a circular orbit is possible there.
However, in the particle distribution given by the steady state these
circular orbits need not necessarily be populated. As a matter of fact we prove
that in a neighborhood of the boundary of the steady state no particles
in the steady state distribution travel on circular orbits.
This neighborhood is in fact large:
Our numerical simulations indicate that the typical range where stars on
circular orbits do not exist is given by $[0.6\,R_b,R_b]$, where $R_b$ denotes
the radius of the steady state. Test particles on circular orbits nevertheless
do in general exist in that region, and their velocity is used to define the
rotation curves.

The Vlasov-Poisson system is given by
\begin{eqnarray}
&\partial_{t}F+v\cdot\nabla_{x}F-\nabla_{x}U\cdot \nabla_{v}F=0,&\\
&\displaystyle\Delta U=4\pi \rho,\ \lim_{|x|\to\infty}U(t,x)=0,&\label{poisson}\\
&\rho(t,x)=\int_{\R^3}F(t,x,v)\,dv.&
\end{eqnarray}
Here $F:\R\times\R^3\times\R^3\to \R_0^+$ is the density function on phase space
of the particle ensemble, i.e., $F=F(t,x,v)$ where $t\in\R$ and $x,v\in\R^3$
denote time, position, and velocity respectively.
The mass of each particle in the ensemble is assumed to be equal
and is normalized to one. The mass density is denoted
by $\rho$ and the gravitational potential by $U$. The latter is given by
\begin{equation}\label{intrep}
U(t,x)=-\int_{\R^3}\frac{\rho(t,y)}{|x-y|}\,dy.
\end{equation}
In this investigation we are interested in extremely flattened
axially symmetric galaxies where all the stars are concentrated
in the $(x_1,x_2)$-plane. We therefore assume that
\[
F(t,x,x_3,v,v_3)=f(t,x,v)\delta(x_3)\delta(v_3),
\]
where from now on $x,v\in\R^2$ and $\delta$ is the Dirac distribution.
The stars in the plane will only experience
a force field parallel to the plane, and the Vlasov-Poisson system for the
density function $f=f(t,x,v), \,x,v\in \R^2,$ takes the form
\begin{eqnarray}
&\partial_{t}f+v\cdot\nabla_{x}f-\nabla_{x}U\cdot\nabla_{v}f=0,&\label{vlasov}\\
&\displaystyle U(t,x)=-\int_{\R^2}\frac{\Sigma(t,y)}{|x-y|}\,dy, &\label{U}\\
&\Sigma(t,x)=\int_{\R^2}f(t,x,v)\,dv;&\label{rho}
\end{eqnarray}
the surface density $\Sigma$ and the spatial density $\rho$ are related through
$\rho(t,x,x_3) = \Sigma(t,x)\delta(x_3)$.
We emphasize that the system (\ref{vlasov})-(\ref{rho}) is not a
two dimensional version of the Vlasov-Poisson system but a special case
of the three dimensional version where the density function is partially
singular.

The study of the Vlasov-Poisson system has a long history in mathematics,
and some of the publications which are relevant here are cited below.
Its history in the astrophysics literature is considerably longer
and to our knowledge started with the investigations of Jeans at the beginning
of the last century. The authors are certainly not
competent to give an appropriate account of the astrophysics
literature on the Vlasov-Poisson system in general.
Some references which are more specific to the issue at hand are
cited below. For general astrophysics background on the Vlasov-Poisson
system we mention the seminal paper of Lynden-Bell (1967)
on violent relaxation, the work of Chavanis (2002) on the thermodynamics
of self-gravitating systems, and the work of Amorisco \& Bertin (2010)
on modelling thin disks in isothermal halos.
More background and references can be found in (Binney \& Tremaine 1987).
The validity of the Vlasov-Poisson system for modelling galaxies
has been challenged in the astrophysics literature, but the
authors take this model as the starting point for the present investigation.

The aim of this investigation is to numerically construct axially
symmetric static solutions of
the system (\ref{vlasov})-(\ref{rho}) as models of disk galaxies, and
to study their rotation curves.
The restriction to static solutions means that $f$ is time independent,
i.e., $f=f(x,v)$, and the restriction to axial symmetry means that
$f$ is invariant under rotations, i.e.,
\begin{equation}
f(Ax,Av)=f(x,v),\, x,v\in\R^2,\, A\in\mbox{SO(2)}.\label{asymm}
\end{equation}
For an axially symmetric system the $x_3$-component or,
using a different notation, the $z$-component
\[
L_z=x_1v_2-x_2v_1
\]
of the particle angular momentum is conserved along particle trajectories.
Since $U$ is time independent the same is true for the particle energy
\begin{equation}\label{E}
E=\frac12 |v|^2+U(x).
\end{equation}
Hence any function of the form
\begin{equation}\label{ansatz}
f(x,v)=\Phi(E,L_z)
\end{equation}
is a solution of the Vlasov equation (\ref{vlasov}).
In fact, for any spherically symmetric, static solution of the
Vlasov-Poisson system $f$ is a function of the particle energy $E$
and the modulus of angular momentum $|x\times v|$.
This is the content of Jeans' theorem,
cf.~(Batt, Faltenbacher \& Horst~1986),
but the corresponding assertion is probably false in the
flat, axially symmetric case.

In the regular three dimensional case the existence theory
for static solutions is well developed,
cf.~(Ramming \& Rein~2013) and the
references there.
In particular, there exists a large class of ansatz functions $\Phi$
which give rise to compactly supported
solutions with finite mass which are stable, cf.~(Rein~2007) and the
references there.
For flat disk solutions, which is the case of interest here,
only a few results are available. Rein~(1999)
and Firt \& Rein~(2006) showed
that solutions with compact support and finite mass exist when
\begin{equation}\label{phit}
\Phi(E)=(E_0-E)_+^{k},\;0<k<1.
\end{equation}
Here $(x)_+=x$ if $x\geq 0$ and $(x)_+=0$ if $x<0$, and $E_0<0$ is a
cut-off energy.
In addition, Firt \& Rein~(2006)
showed that the steady states are stable against flat
perturbations.
We note that the ansatz (\ref{phit}) is independent of $L_z$,
but for the purpose of our present investigation it is crucial that the
ansatz admits a dependence on $L_z$.

Let us mention a few of the previous studies on
models of disk solutions where the aim is to obtain rotation curves which
agree with observations. Pedraza, Ramos-Caro \& Gonz\'{a}lez (2008)
study a family of axisymmetric flat models of generalized
Kalnajs type which give satisfactory behavior of the rotational curves
without the assumption of a dark matter halo.
Gonz\'{a}lez, Plata \& Ramos-Caro~(2010) start from the
observational data of four galaxies and construct models with the
corresponding densities and potentials.
The effects of relativistic corrections are studied in
(Ramos-Caro, Agn \& Pedraza~2012, Nguyen \& Lingam~2013), and a
model motivated by renormalization group corrections is investigated
in (Rodrigues, Letelier \& Shapiro~2010).
Rotation curves obtained in alternative gravitational
theories such as Modified Newtonian Dynamics (MOND) and
Scalar-Tensor-Vector Gravity (STVG) can be found
in~(Famaey \& McGaugh~2012) and (Moffat 2006),
respectively.
Rein~(2013) studies
steady states of a MONDian Vlasov-Poisson system.

The outline of the present paper is as follows. In the next section we utilize
the symmetry assumption and derive the system of equations which we solve
numerically. In Section 3 the equation for circular orbits is discussed.
In particular we show that
for spherically symmetric and flat axially symmetric steady states
there are no stars on circular orbits in the neighborhood of the boundary of
the steady state.  In Section 4 the form of the ansatz function
which we consider is given and the ingoing parameters are discussed.
The numerical algorithm is also briefly analyzed. Our numerical results are
presented in Section 5. We find a large class of solutions which give rise
to approximately flat rotation curves as well as rotation curves which are
slightly decreasing or increasing. The range where stars in circular orbits
do not exist is computed numerically in a couple of cases and an estimate of
the additional mass required to obtain the rotation velocities
using a Keplerian approach is given. In the last section we compare the
predictions of our models with some measurements from real galaxies.

\section{A reduced system of  equations}
\setcounter{equation}{0}
In this section we derive a simplified form of the flat, static,
axially symmetric
Vlasov-Poisson system (\ref{vlasov})-(\ref{rho}) by using the ansatz
(\ref{ansatz}) and the symmetry assumption (\ref{asymm}).

For axially symmetric steady states the mass density $\Sigma$ and the potential
$U$ are functions of $r:=\sqrt{x_1^2+x_2^2}$.
In view of (\ref{ansatz}) and (\ref{rho})
the change of variables $(v_1,v_2) \mapsto (E,L_z)$ implies that
\begin{equation}\label{rhonoE0}
\Sigma(r)=2\int_{U(r)}^{\infty}\int_{-\sqrt{2r^2(E-U(r))}}^{\sqrt{2r^2(E-U(r))}}
\frac{\Phi(E,L_z)\,dL_z\,dE}{\sqrt{2r^2(E-U(r))-L_z^2}}.
\end{equation}
Next we adapt the formula for the potential $U$
to the case of axial symmetry.
We introduce polar coordinates so that $x=(r\cos\varphi, r\sin\varphi)$.
In view of (\ref{asymm}), (\ref{U}) and (\ref{rho}) it follows that
\[
U(x)=U(r\cos\varphi, r\sin\varphi)=U(r,0)=:U(r).
\]
Denoting $y=(s\cos\theta,s\sin\theta)$,
\[
|(r,0)-y|^2 =  r^2-2rs\cos\theta +s^2
=  (r+s)^2\Big[1-k^2\cos^2\Big(\frac{\theta}{2}\Big)\Big],
\]
where
\[
k=\frac{2\sqrt{rs}}{r+s}.
\]
Using the complete elliptic integral of the first kind
\[
K(\xi)=\int_0^{\pi/2}\frac{d\theta}{\sqrt{1-\xi^2\sin^2\theta}}
=\int_0^1\frac{dt}{\sqrt{1-\xi^2 t^2}\sqrt{1-t^2}},\quad 0\le\xi<1,
\]
it follows upon substituting
$t=\cos(\frac{\theta}{2})$ that
\begin{eqnarray}
U(r)
& = &
-\int_0^\infty \int_0^{2\pi} \frac{s\Sigma(s)}{|(r,0)-y|}d\theta\,ds\nonumber\\
& = &
-\int_0^\infty \frac{s\Sigma(s)}{(r+s)}
\int_0^{2\pi}\frac{d\theta}{\sqrt{1-k^2\cos^2(\frac{\theta}{2})}} ds\nonumber\\
& = &
-\,4\int_0^\infty \frac{s\Sigma(s)}{r+s}\,K(k)\,ds.\label{UK}
\end{eqnarray}
The equations (\ref{rhonoE0}) and (\ref{UK}) constitute the system we use
to numerically construct axially symmetric flat solutions of the
Vlasov-Poisson system. However, we need to be reasonably certain that this
leads to steady states which have finite total mass
and compact support.
In the spherically symmetric case a necessary condition
for these physically desirable properties of the resulting steady states
is that there is a cut-off
energy such that the distribution function vanishes for
sufficiently large particle energies, cf.\
(Rein \& Rendall~2000).
We now show that this condition is
quite natural or necessary also in the flat, axially symmetric case.

\begin{proposition} \label{cutoffex}
Assume that $(f,\Sigma,U)$ is a steady state of the flat, axially symmetric
Vlasov-Poisson system in the sense that
$f=\Phi(E,L_z)$ for some measurable function $\Phi : \R^2 \to [0,\infty[$,
that (\ref{rhonoE0}) defines a measurable function
and (\ref{UK}) holds. Assume further that
the steady state has finite mass
\[
M= 2\pi \int_0^\infty r \Sigma(r)\, dr < \infty
\]
and $\lim_{r\to \infty} U(r)=0$.
Then $\Phi(E,L_z)=0$ for almost all $E > 0$
and $L_z \in \R$.
\end{proposition}

\begin{remark}
A potential given by (\ref{UK}) satisfies the boundary condition
$\lim_{r\to \infty} U(r)=0$ at least formally, and also rigorously
provided the steady state has finite mass and is properly isolated.
\end{remark}

\noindent
\textit{Proof of Proposition~\ref{cutoffex}.}
By assumption there exist $r_0>0$ and $u_0 > 0$ such that
\[
-u_0 \leq U(r) \leq 0\ \mbox{for}\ r\geq r_0.
\]
Combining the formula for total mass with (\ref{rhonoE0}) implies that
\begin{eqnarray*}
M
&=&
4 \pi \int_0^\infty \int_{U(r)}^\infty
\int_{-\sqrt{2r^2(E-U(r))}}^{\sqrt{2r^2(E-U(r))}}
\frac{\Phi(E,L_z)}{\sqrt{2r^2 (E-U(r))-L_z^2}}\,dL_z\,dE\, r \, dr\\
&\geq&
4 \pi \int_{r_0}^\infty \int_0^\infty
\int_{-\sqrt{2r^2 E}}^{\sqrt{2r^2 E}}
\frac{\Phi(E,L_z)}{\sqrt{2r^2(E+u_0)}}\,dL_z\,dE\, r \, dr\\
&=&
2\sqrt{2} \pi \int_0^\infty \int_{-\infty}^\infty
\frac{\Phi(E,L_z)}{\sqrt{E+u_0}}
\int_{\max\left(r_0,|L_z|/\sqrt{2 E}\right)}^\infty dr\, \,dL_z\,dE.
\end{eqnarray*}
Since the $r$ integral is infinite this implies that $\Phi$
vanishes almost everywhere on $]0,\infty[\times \R$ as claimed.
\prfe

Motivated by the discussion above we from now on assume
that the ansatz functions $\Phi$ which
we consider have the property that
\begin{equation}\label{cutoff}
\Phi(E,L_z)=0\ \mbox{ if }\ E\geq E_0
\end{equation}
for a given cut-off energy $E_0$. Together with (\ref{rhonoE0})
this implies that
\begin{equation}\label{rhoE0}
\Sigma(r)=2\int_{U(r)}^{E_0}\int_{-\sqrt{2r^2(E-U(r))}}^{\sqrt{2r^2(E-U(r))}}
\frac{\Phi(E,L_z)\,dL_z\,dE}{\sqrt{2r^2(E-U(r))-L_z^2}}\ \mbox{where}\ U(r)<E_0,
\end{equation}
and $\Sigma(r)=0$ where $U(r)\geq E_0$.
The exact form of the ansatz functions $\Phi$
which we study is given in Section~4.
\section{Circular orbits}
\setcounter{equation}{0}
The equation for circular orbits is standard, cf.\
(Binney \& Tremaine~1987),
but let us here relate it to the Vlasov equation.
Let the radial velocity be denoted by $w$, i.e., $w=x\cdot v /r$.
The coordinate transformation $(v_1,v_2)\mapsto (w,L_z)$, and the fact
that the angular momentum $L_z$ is conserved along along particle trajectories
implies that the radius $r(s)$ and
the radial velocity $w(s)$ along a trajectory is given by
\begin{eqnarray*}
\frac{d r}{ds}&=& w,\\
\frac{d w}{ds}&=&\frac{L_z^2}{r^3}-U'(r).
\end{eqnarray*}
The radial velocity $w$ is zero along a circular orbit and in
particular we have $dw/ds=0$ which implies that
\[
\frac{L_z^2}{r^3}-U'(r)=0.
\]
The circular velocity $v_c$ is given by $v_c=L_z/r$, and the equation
for a circular orbit is thus given by
\begin{equation}\label{circorbit}
v_c^2=r\, U'(r).
\end{equation}
We note that $U'(r)$ has to be positive for a circular orbit of radius $r$ to exist.
In the spherically symmetric situation this is always the case.
However, in the present case this is in general not true.
It is straightforward to construct axially symmetric mass densities $\Sigma$
such that $U'(r)<0$ for some $r>0$.
Moreover, below we will find self-consistent solutions of the
Vlasov-Poisson system such that $U'(r)<0$ on some interval of the radius $r$,
cf.~Figures~\ref{fig8} and~\ref{fig9}. More interestingly, it turns out
in the particle ensemble given by the density $f$ with
an ansatz function which satisfies (\ref{cutoff}) no particles exist
on circular orbits in a neighborhood of the boundary of the steady state.

Let $R_b$ denote the boundary of the steady state. We have the following result
which we state and prove in the flat, axially symmetric case as well as in
the spherically symmetric case.
\begin{theorem}\label{nocircorbits}
Consider a non-trivial compactly supported steady state of
the flat, axially symmetric Vlasov-Poisson system or the spherically
symmetric Vlasov-Poisson system
with the property (\ref{cutoff}). Then there is an $\epsilon>0$ such that
for $r\in [R_b-\epsilon,R_b]$ no
circular orbits exist in the particle distribution given by $f$.
\end{theorem}
\begin{remark}
\begin{itemize}
\item[(a)]
Numerically we find that in the flat case and for our ansatz
functions the typical interval where
there are no stars in circular orbits
is approximately given by $[0.6 R_b,R_b]$.
\item[(b)]
It should be stressed that if $U'(r)>0$ for some radius $r>0$
then test particles with the proper circular velocity do travel on the circle of
radius $r$. The result above only says that {\em the particle distribution
given by the steady state} does not contain such particles, i.e., stars.
We nevertheless compute the rotation curves below by using equation
(\ref{circorbit}) and compare these with observational data.
Hence the interpretation of our rotation curves is that they correspond
to the circular orbits of test particles in the gravitational
field generated by the steady state of the Vlasov-Poisson system.
In measurements of the rotation curves of real galaxies it is to our knowledge
the orbital velocity of gas particles which is measured.
Since the mass of a gas particle is small compared to the mass of a star,
the gas particle can be treated as a test particle in the gravitational
field generated by the stars in the galaxy. It is an interesting problem to
construct self-consistent steady states where two types of
particles---stars and gas particles---are present and to compute rotation curves
from the distribution of the gas particles. In any case we believe that these
different viewpoints in defining and interpreting rotational velocities and
rotation curves for galaxy models should be relevant also for astrophysical
applications.
\end{itemize}
\end{remark}

\noindent
\textit{Proof of Theorem~\ref{nocircorbits}.}
We first consider the spherically symmetric case.
The modulus of the angular momentum $L$ is then given by $L=|x\times v|$.
Analogously to the derivation above we have that for a
particle on a circular orbit
\[
\frac{L^2}{r^3} - U'(r) = 0.
\]
Since $U$ is spherically symmetric, $U'(r)=m(r)/r^2$ where
\[
m(r)= 4\pi \int_0^r s^2\Sigma(s)\,ds.
\]
Hence, on a circular orbit $L = r m(r).$
The corresponding energy for such a particle is given by
\[
E = \frac12 \frac{m(r)}{r}+U(r).
\]
At the boundary of the steady state we have in view of
(\ref{cutoff}) and (\ref{rhoE0}) that $U(R_b)=E_0$. Moreover,
in spherical symmetry the potential $U$ is given by
\[
U(r)=-4\pi\frac{1}{r}\int_0^r  s^2 \Sigma(s)\,ds-4\pi\int_r^{\infty}  s \Sigma(s)\,ds,
\]
so that $U(R_b)=-M/R_b,$ where $M$ is the total mass given by
\[
M=4\pi \int_0^{R_b}  s^2\Sigma(s)\,ds.
\]
Note that $M>0$ since the steady state is non-trivial.
Hence, as $r\to R_b$ the particle energy
\[
E\to -\frac{M}{2R_b}>-\frac{M}{R_b}=E_0.
\]
In a neighborhood of the boundary it thus follows that particles
on circular orbits must have a particle energy $E$ which is larger
than $E_0$. By the cut-off condition (\ref{cutoff})
no such particles exist in the particle distribution
of the steady state.
Let us turn to the axially symmetric flat case.
On p.~267 in~(Dejonghe~1986)
the following form of the potential $U$ is derived:
\[
U(r)=-\frac{4}{r}\int_0^r s\Sigma(s)\,K\left(\frac{s}{r}\right)\,ds
-4\int_r^{\infty}\Sigma(s)\,K\left(\frac{r}{s}\right)\,ds.
\]
Using the relation
\[
K'(k)=\frac{F(k)}{k(1-k^2)}-\frac{K(k)}{k},
\]
where $F$ is the complete elliptic integral of the second kind
(which usually is denoted by $E$),
it follows by a straightforward computation that
\begin{equation}\label{UprimeHD}
U'(R_b)=\frac{4}{R_b}\int_0^{R_b}s\Sigma(s)
\frac{F(\frac{s}{R_b})}{1-\frac{s^2}{R_b^2}}\,ds,
\end{equation}
since $\Sigma(r)=0$ for $r \geq R_b$.
On a circular orbit it holds that
\[
\frac{L_z^2}{r^3}=U'(r),
\]
and the particle energy on such an orbit is thus given by
\[
E=\frac12 r\,U'(r)+U(r).
\]
Hence
\[
E\to \frac12 R_b\,U'(R_b)+E_0 \mbox{ as }r\to R_b,
\]
since $U(R_b)=E_0$ in view of (\ref{cutoff}).
Now the first term on the right hand side is strictly positive for a
non-trivial steady state in view of (\ref{UprimeHD}) since $F\geq 1$.
By continuity it follows that $E$ is larger than $E_0$ in a neighborhood
of the boundary.
\prfe

\section{The ansatz and the numerical algorithm}
\setcounter{equation}{0}
We consider the following ansatz function:
\begin{equation}
\Phi(E,L_z) = A\,(E_0-E)_+^k\,(1-Q\,|L_z|)_+^l. \label{ansatz1}
\end{equation}
Here $A>0$, $E_0<0$, $Q\geq 0$, $k\geq 0$, and $l$ are constants.
An alternative ansatz is given by
\[
\Phi(E,L_z)=A\,(E_0-E)_+^k\,(1-Q\,L_z)_+^l\,H(L_z),
\]
where $H$ is the Heaviside function.
This may be introduced to ensure that all particles move in the same direction
about the axis of symmetry so that the solution has non-vanishing total angular
momentum, i.e., the disk rotates. However, by fixing the mass these two
versions of the ansatz function result in the same density-potential pair and
in particular the same rotation curve.

The constant $A$ is merely a normalization constant which controls the total
mass of the solution. Hence, when a solution is depicted we give its total
mass rather than the value of $A$. In this context we point out that the
ansatz (\ref{ansatz1}) can be written as
\begin{equation}
\Phi(E,L_z)=\tilde{A}\,(E_0-E)_+^k\,(L_0-|L_z|)_+^l,\label{ansatz2}
\end{equation}
where $\tilde{A}=A\,Q^l$ and $L_0=1/Q$ is the cut-off for the angular momentum
$L_z$. In principle this form seems more natural than the form given in
(\ref{ansatz1}). However, an important test case for our numerical algorithm
is when $Q=0$ since the ansatz is then independent on $L_z$,  and the
integration in $L_z$ in (\ref{rhoE0}) can be carried out explicitly.
This is one reason why we use (\ref{ansatz1}) rather than (\ref{ansatz2}).

\begin{remark}
\begin{itemize}
\item[(a)]
The ansatz~(\ref{ansatz1}) has the property that $\Phi$ is decreasing as a
function of $E$ for fixed $L_z$. In the regular three dimensional case this
property is well-known to be essential for stability,
cf.\ (Rein~2007). The
ansatz for the Kalnajs disk, cf.\
(Binney \& Tremaine~1987), is on the other hand
\textit{increasing} in $E$ which indicates that these solutions are unstable,
which is also supported by the results of Kalnajs (1972).
\item[(b)]
The following ansatz
\[
\Phi(E,L_z)=A\,(E_0-E)_+^k\,(\epsilon+(1-Q\,|L_z|)_+^l)
\]
gives roughly the same results when $\epsilon>0$ is small. The reason for
introducing the constant $\epsilon$ in the ansatz is the following.
The method in
(Rein~1999) for showing the existence of stable, compactly
supported
solutions with finite mass does in fact apply when $\Phi$ also depends on $L_z$.
However, it then requires that the factor in~(\ref{ansatz1}) which depends on
$L_z$ is bounded from below by a positive constant. This property holds if $\epsilon>0$.
It should on the other hand be pointed out that the method in (Rein~1999)
constrains the value of $k$ to $1/2<k<1$ when a dependence on $L_z$ is
admitted. The cases which give rise to flat rotation curves in our numerical
simulations require that $k$ is small, in particular that $k<1/2$. Hence, it
is an open and important problem to show existence of compactly supported
solutions with finite mass for ansatz functions of type (\ref{ansatz1})
when $k<1/2$.
\end{itemize}
\end{remark}

In the numerical simulations below we choose $k=0$ for simplicity, since the
results are not affected in an essential way as long as this value is small.
For the present ansatz, the cut-off energy $E_0$ determines the extent of the
support of the solution, although there is in general no guarantee that the
solutions have finite extent even if the cut-off condition is satisfied.
Nevertheless, in our case the influence of the parameter $E_0$ is of
limited interest and will not affect the qualitative behavior of the
solutions. Hence, for our purposes the essential parameters are $l$
and $Q$, and below we mainly study their influence on the solutions.


The system of equations (\ref{UK}), (\ref{rhoE0}) is solved by an
iteration scheme of the following type:
\begin{itemize}
\item
Choose some start-up density $\Sigma_0$ which is non-negative and has a
prescribed mass $M$.
\item
Compute the potential $U_0$ induced by $\Sigma_0$
via (\ref{UK}); the complete elliptic integral appearing there is computed
using the gsl package.
\item
Compute the spatial density $\tilde \Sigma$ from (\ref{rhoE0}).
\item
Define $\Sigma_1 = c \tilde \Sigma$ where $c$ is chosen such that $\Sigma_1$
again has mass $M$.
\item
Return to the first step with $\Sigma_0:=\Sigma_1$.
\end{itemize}
In order to obtain convergence it is crucial
that in each iteration the mass of $\Sigma$ is kept constant.
It is easy to see that a fixed-point of this iteration is a steady state
of the desired form where the constant $A$ in the ansatz function has been
rescaled. We emphasize that in all cases studied for the ansatz
(\ref{ansatz1}) convergence is obtained in the sense that the difference
between two consecutive iterates can be made as small as we wish by improving
the resolution. Moreover, even if the given initial iterate is very different
from the solution, the iteration sequence quickly starts to converge. As
opposed to this, using the ansatz for the Kalnajs disk as given in
(Binney \& Tremaine~1987)
we obtain convergence only for initial iterates $\Sigma_0$
which are very close to the
solution. Since the Kalnajs disks are expected to be unstable, and since
the convergence in our numerical algorithm is very hard to achieve in this
case, we consider the convincing convergence of our algorithm for the
ansatz (\ref{ansatz1}) to be an indication of the stability
of these solutions.

The complete elliptic integral $K$ becomes singular when its argument approaches unity,
i.e., when $s=r$ in \eqref{UK}. Since this equation involves only a one-dimensional
integration we can handle this fact by using a sufficiently high resolution
in the radial variable and replacing the cases $s=r$ by $s=r+dr$.
In a fully three dimensional, axially symmetric situation more
sophisticated methods will be required as discussed in (Hur\'{e} 2005).

\section{The numerical results}
\begin{figure}
\centering
\begin{subfigure}{.5\textwidth}
  \centering
  \includegraphics[width=1.0\linewidth]{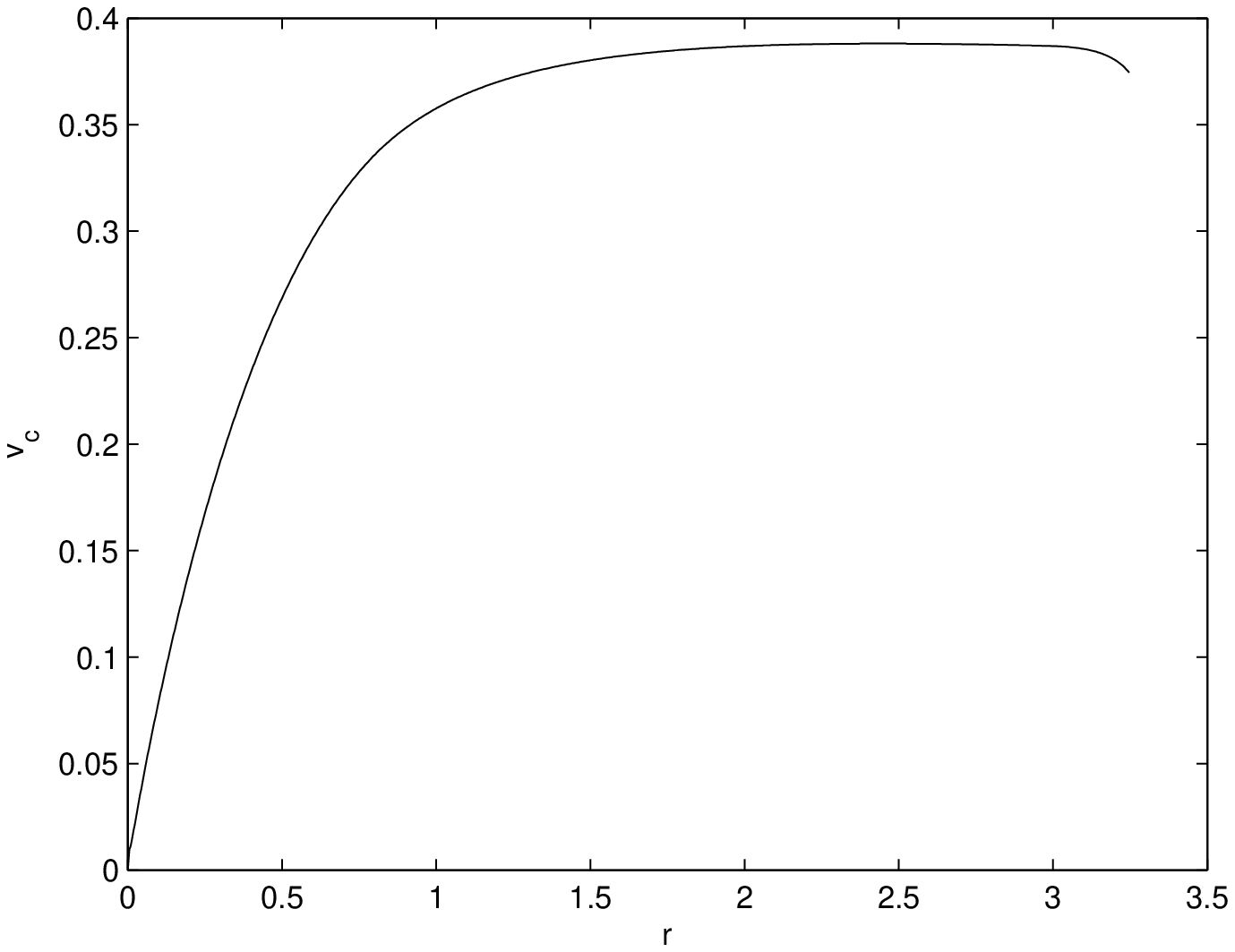}
  \caption{Rotation velocity $v_c$ versus radius $r$}
  \label{fig1:sub1}
\end{subfigure}%
\begin{subfigure}{.5\textwidth}
  \centering
  \includegraphics[width=1.0\linewidth]{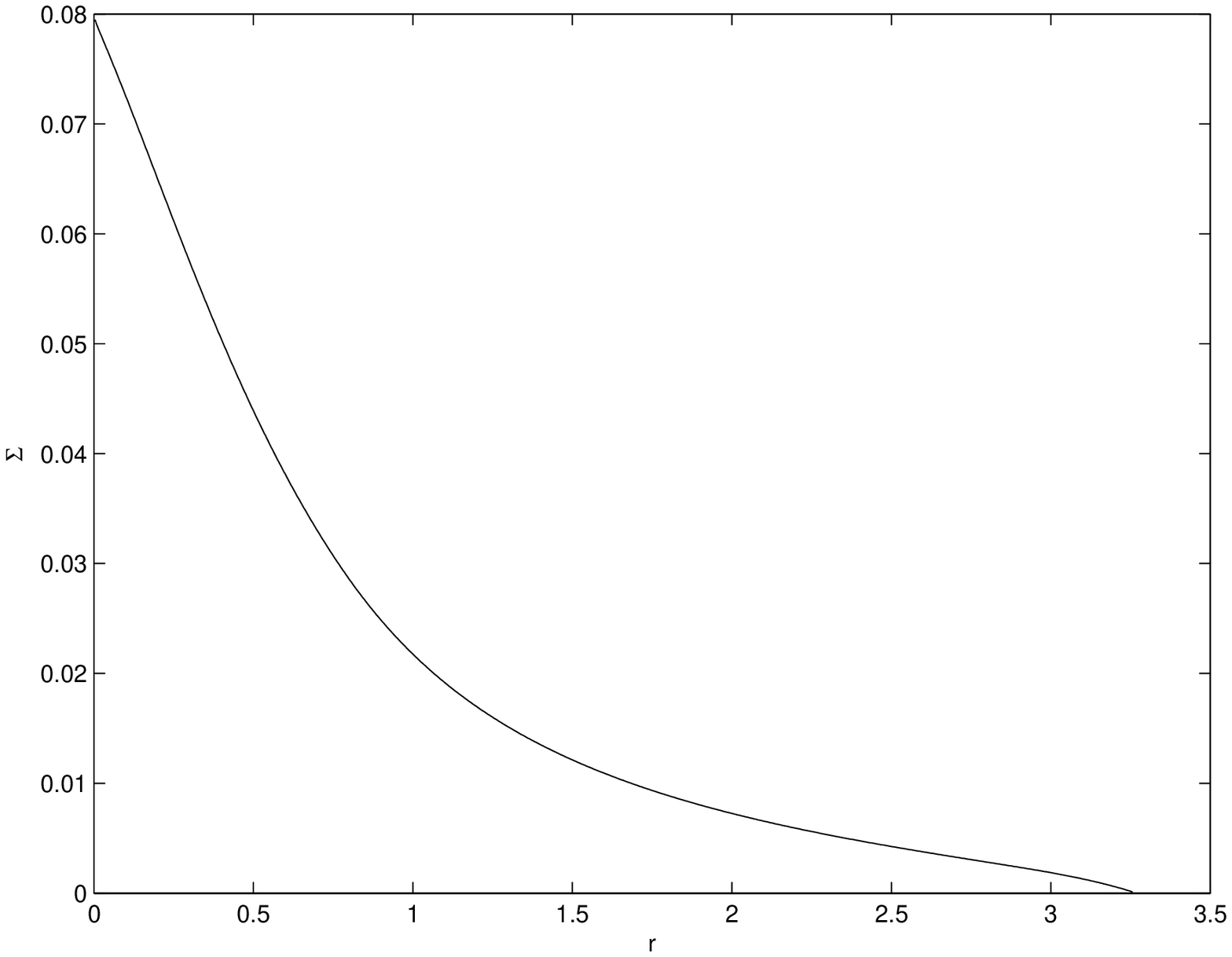}
  \caption{Density $\Sigma$ versus radius $r$}
  \label{fig1:sub2}
\end{subfigure}
\caption{$M=0.3,\;E_0=-0.1,\; l=1.0,\; Q=2.0$}
\label{fig1}
\end{figure}

\begin{figure}[htbp]
\begin{center}
\scalebox{.50}{\includegraphics{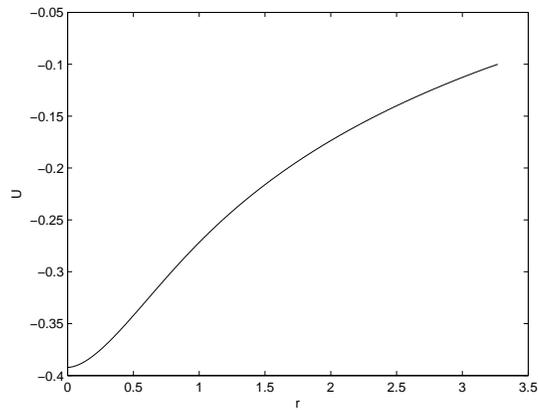}}
\end{center}
\caption{$U$ versus radius $r$ for the steady state in Figure~\ref{fig1}}
\label{fig2}
\end{figure}

\begin{figure}
\centering
\begin{subfigure}{.5\textwidth}
  \centering
  \includegraphics[width=1.0\linewidth]{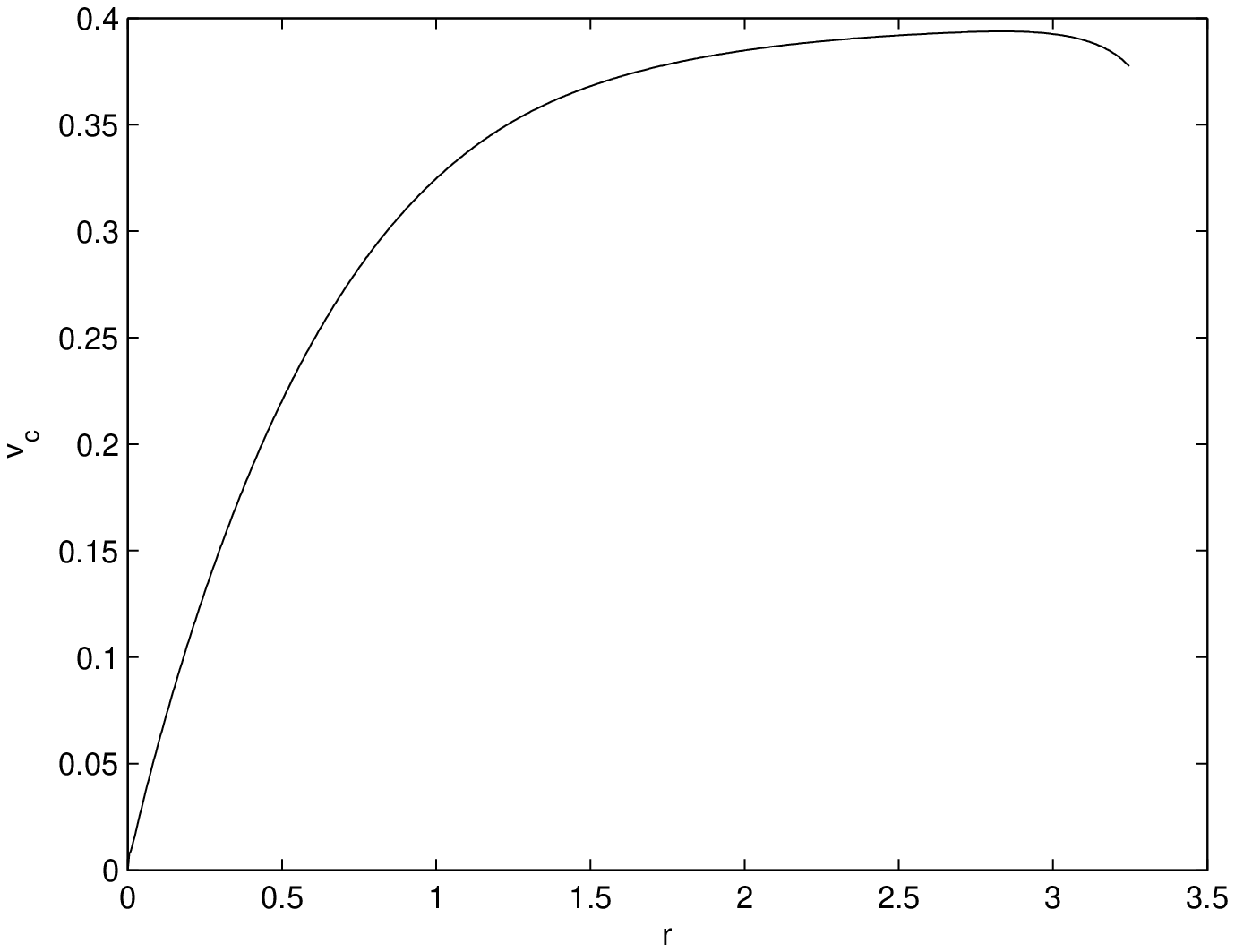}
  \caption{Rotation velocity $v_c$ versus radius $r$}
  \label{fig3:sub1}
\end{subfigure}%
\begin{subfigure}{.5\textwidth}
  \centering
  \includegraphics[width=1.0\linewidth]{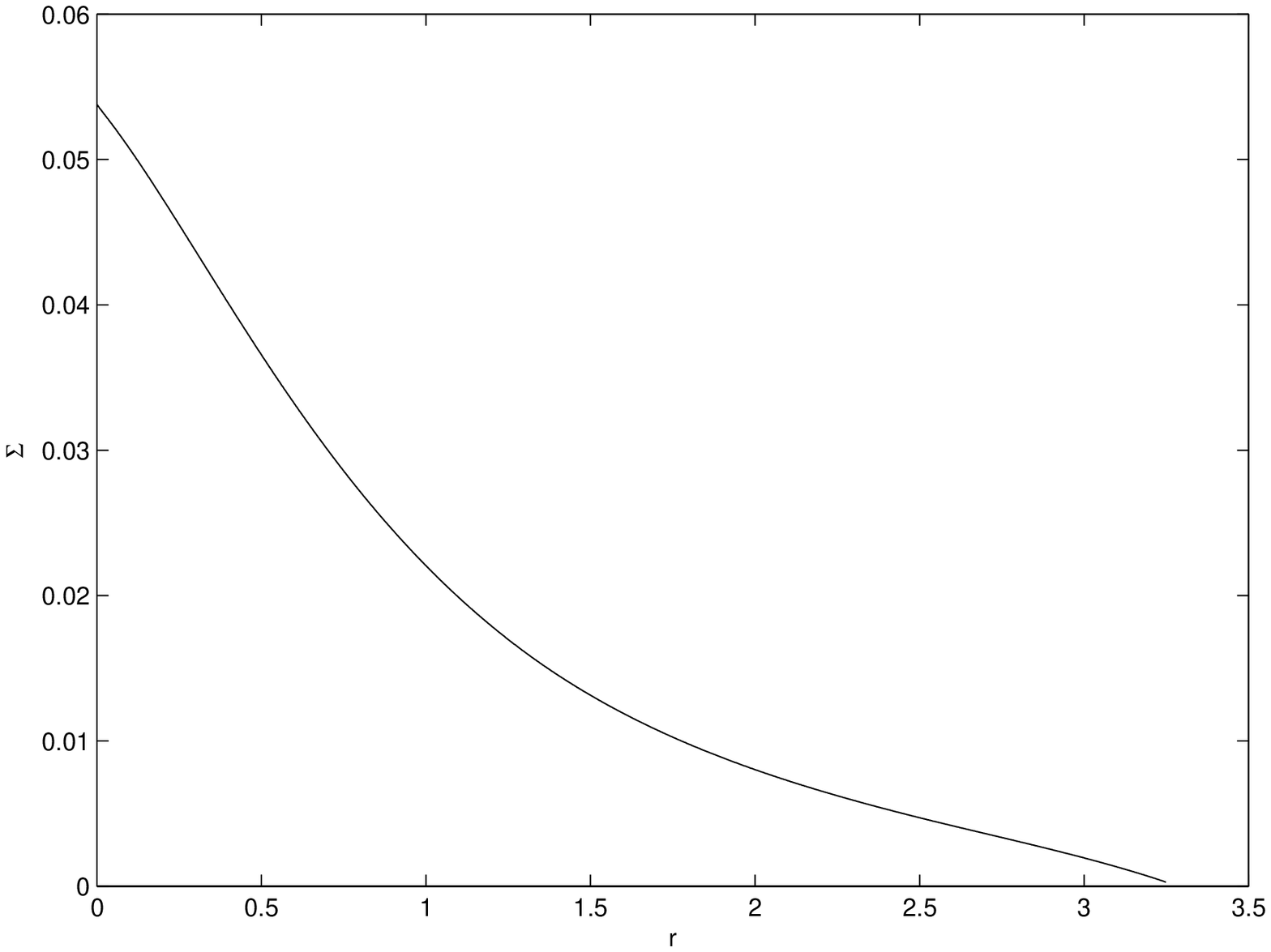}
  \caption{Density $\Sigma$ versus radius $r$}
  \label{fig3:sub2}
\end{subfigure}
\caption{$M=0.3,\;E_0=-0.1,\; l=1.0,\; Q=1.5$}
\label{fig3}
\end{figure}

\begin{figure}
\centering
\begin{subfigure}{.5\textwidth}
  \centering
  \includegraphics[width=1.0\linewidth]{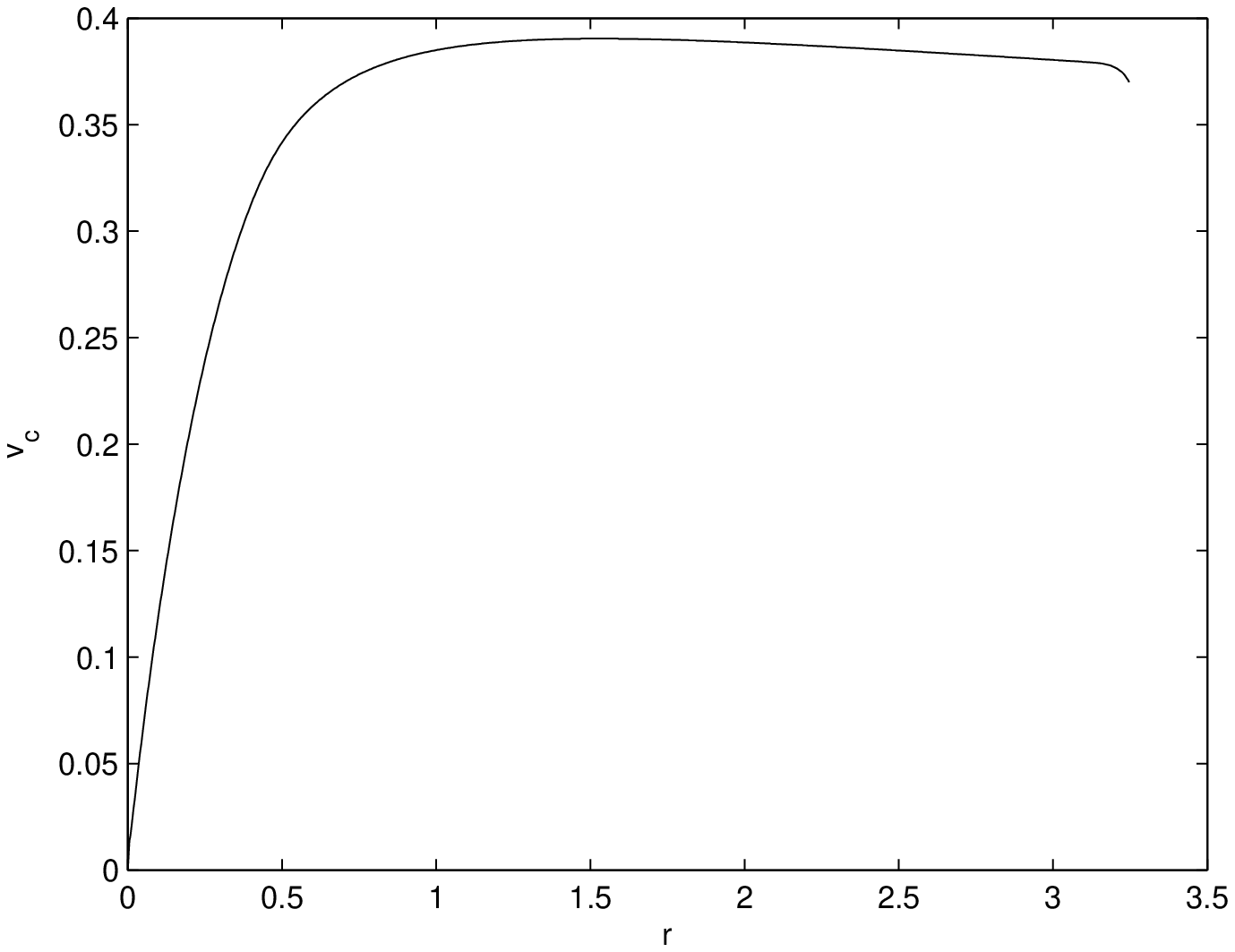}
  \caption{Rotation velocity $v_c$ versus radius $r$}
  \label{fig4:sub1}
\end{subfigure}%
\begin{subfigure}{.5\textwidth}
  \centering
  \includegraphics[width=1.0\linewidth]{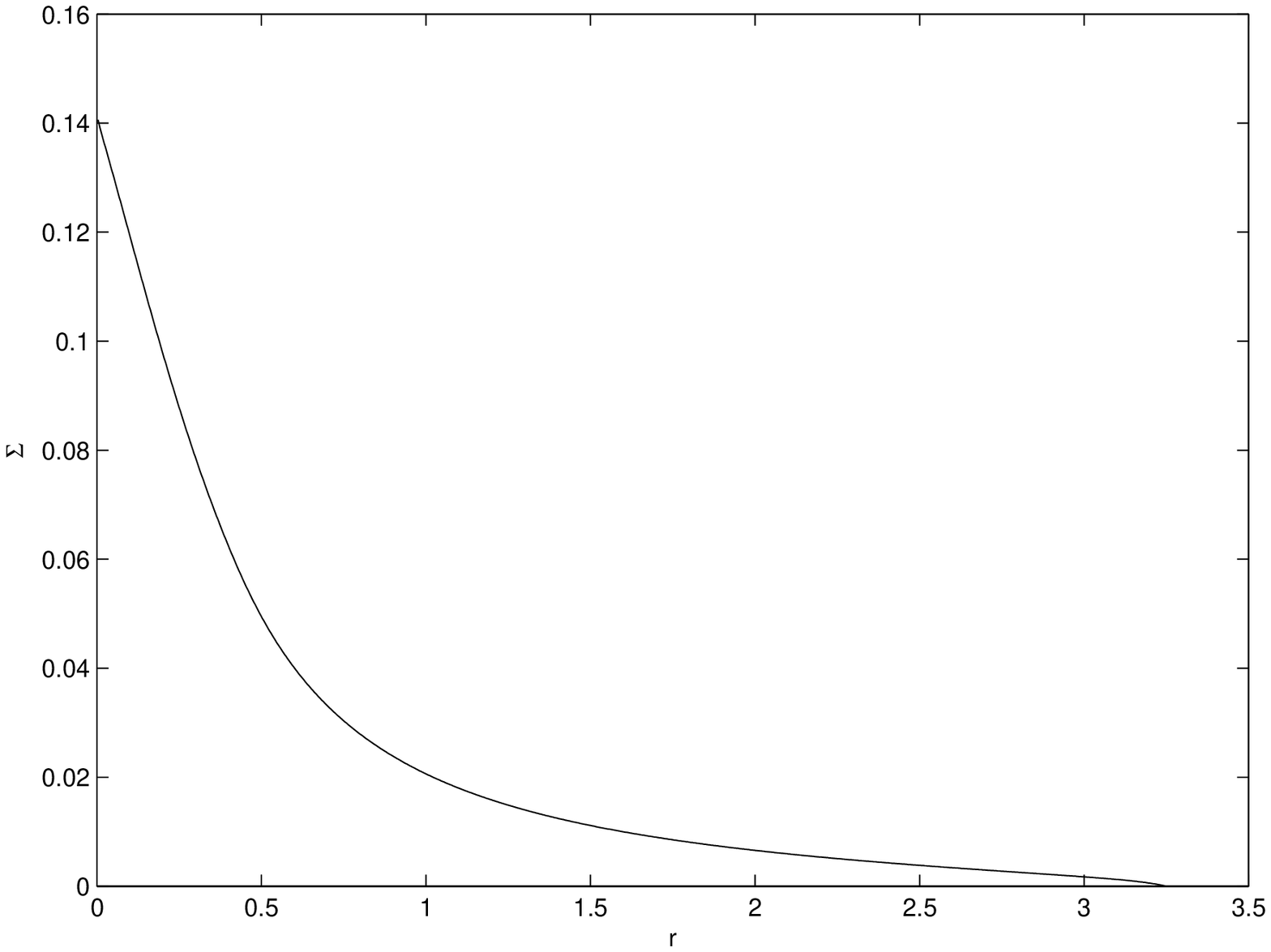}
  \caption{Density $\Sigma$ versus radius $r$}
  \label{fig4:sub2}
\end{subfigure}
\caption{$M=0.3,\;E_0=-0.1,\; l=1.0,\; Q=3.0$}
\label{fig4}
\end{figure}

\setcounter{equation}{0}
As mentioned in the previous section, the important parameters in this
investigation are $l$ and $Q$. In Figure~\ref{fig1:sub1} the rotation
velocity $v_c$  versus the radius $r$ is depicted
in the case $l=1$ and $Q=2$, and in Figure~\ref{fig1:sub2} the corresponding
mass density is shown.
We notice that the rotation curve is approximately flat for values of $r$
reaching almost all the way to the boundary of the steady state.
The boundary is situated where the mass density vanishes.
The potential $U$ for this steady state is depicted in Figure~\ref{fig2}.
We note that $U=E_0$ at the boundary of the steady state. In most cases we
omit the potential $U$, since its features are quite similar in the situations
we study, but important exceptions are given below where $U$ is not monotone.
In Figure~\ref{fig3} and Figure~\ref{fig4} we show the behavior of the
rotational velocity and the mass density when the parameter $Q$ is varied
for the case in Figure~\ref{fig1}.
We notice that the rotation velocity can be slightly increasing or slightly
decreasing in a large fraction of the support. It is of course interesting to
compare the shape of the rotation curves to data from observations,
cf.\ e.g.\
(Fuchs~2001, Verheijen \& Sancisi~2001, Gentile et al~2004, Salucci~2007,
Blok et al~2008, Roos~2010).
We find that the shape of the rotation curves in
Figures~\ref{fig1}, \ref{fig3}, and \ref{fig4} agrees nicely with the
shape of the rotation curves obtained in observations. In the next
section we give in fact examples from real galaxies and we find
solutions that match the observations very well.

\begin{figure}
\centering
\begin{subfigure}{.5\textwidth}
  \centering
  \includegraphics[width=1.0\linewidth]{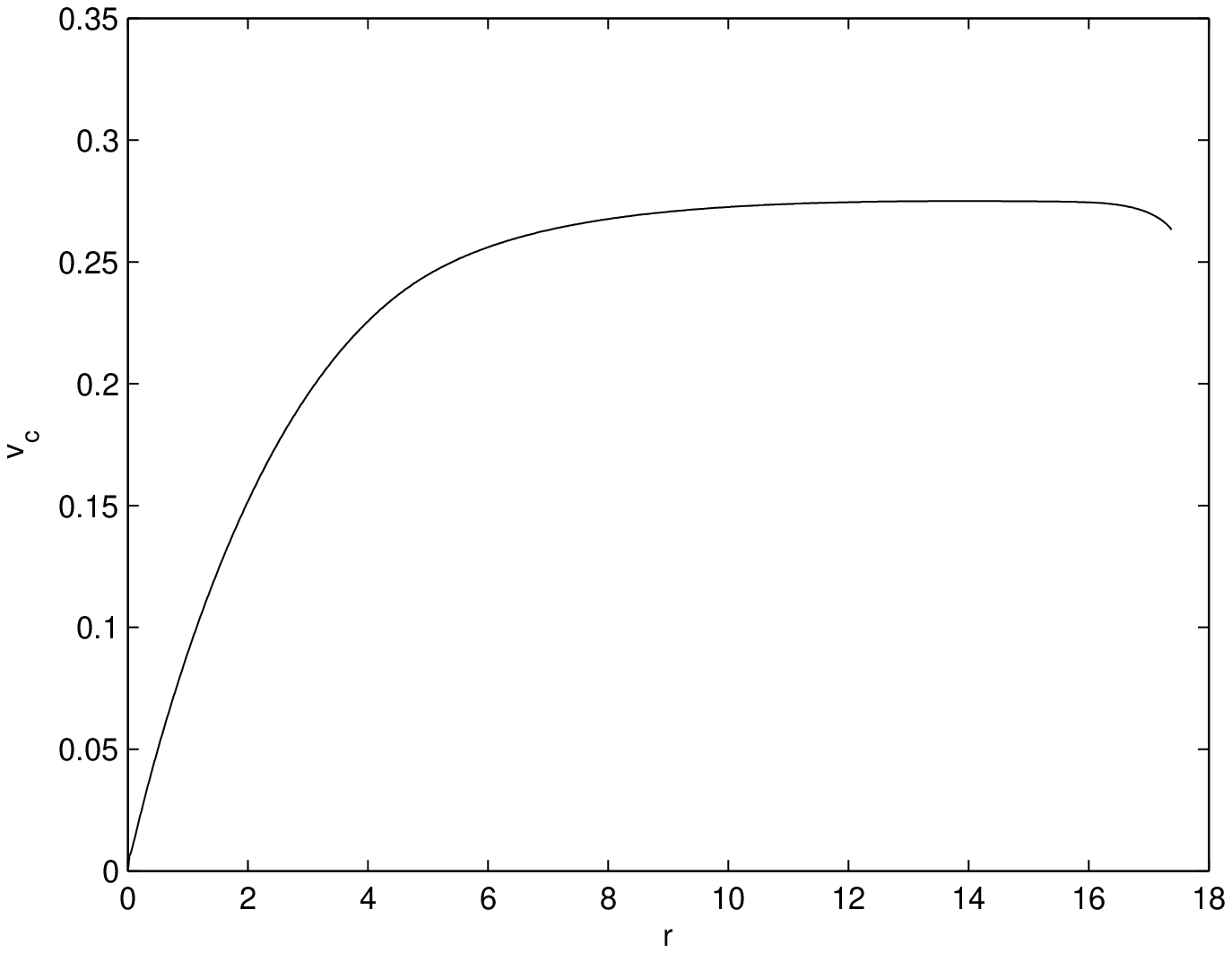}
  \caption{Rotation velocity $v_c$ versus radius $r$}
  \label{fig5:sub1}
\end{subfigure}%
\begin{subfigure}{.5\textwidth}
  \centering
  \includegraphics[width=1.0\linewidth]{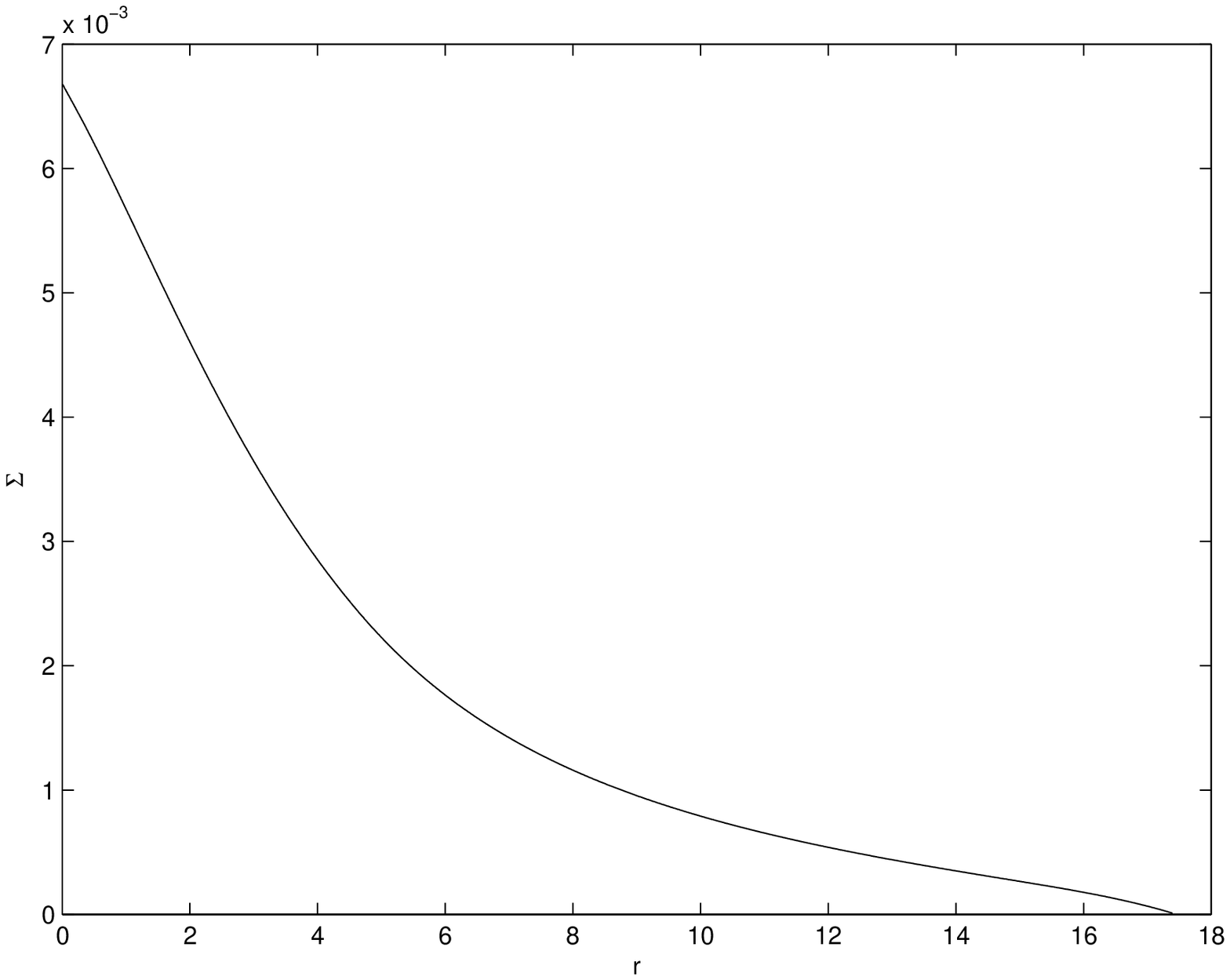}
  \caption{Density $\Sigma$ versus radius $r$}
  \label{fig5:sub2}
\end{subfigure}
\caption{$M=0.8,\;E_0=-0.05,\; l=1.0,\; Q=0.5$}
\label{fig5}
\end{figure}

Before studying the effect of changing the parameter $l$ we note that
the parameters $M$ (reflected by the choice of $A$) and $E_0$ affect
the support and amplitude of $v_c$ and $\Sigma$. In Figure~\ref{fig5}
we again obtain an approximately flat rotation curve,
but in this case the support extends further out and the mass is larger.
We point out that this change does effect the slope of the curve if $Q$
is kept unchanged but in Figure~\ref{fig5} $Q$ has been modified to $Q=0.5$.

\begin{figure}
\centering
\begin{subfigure}{.5\textwidth}
  \centering
  \includegraphics[width=1.0\linewidth]{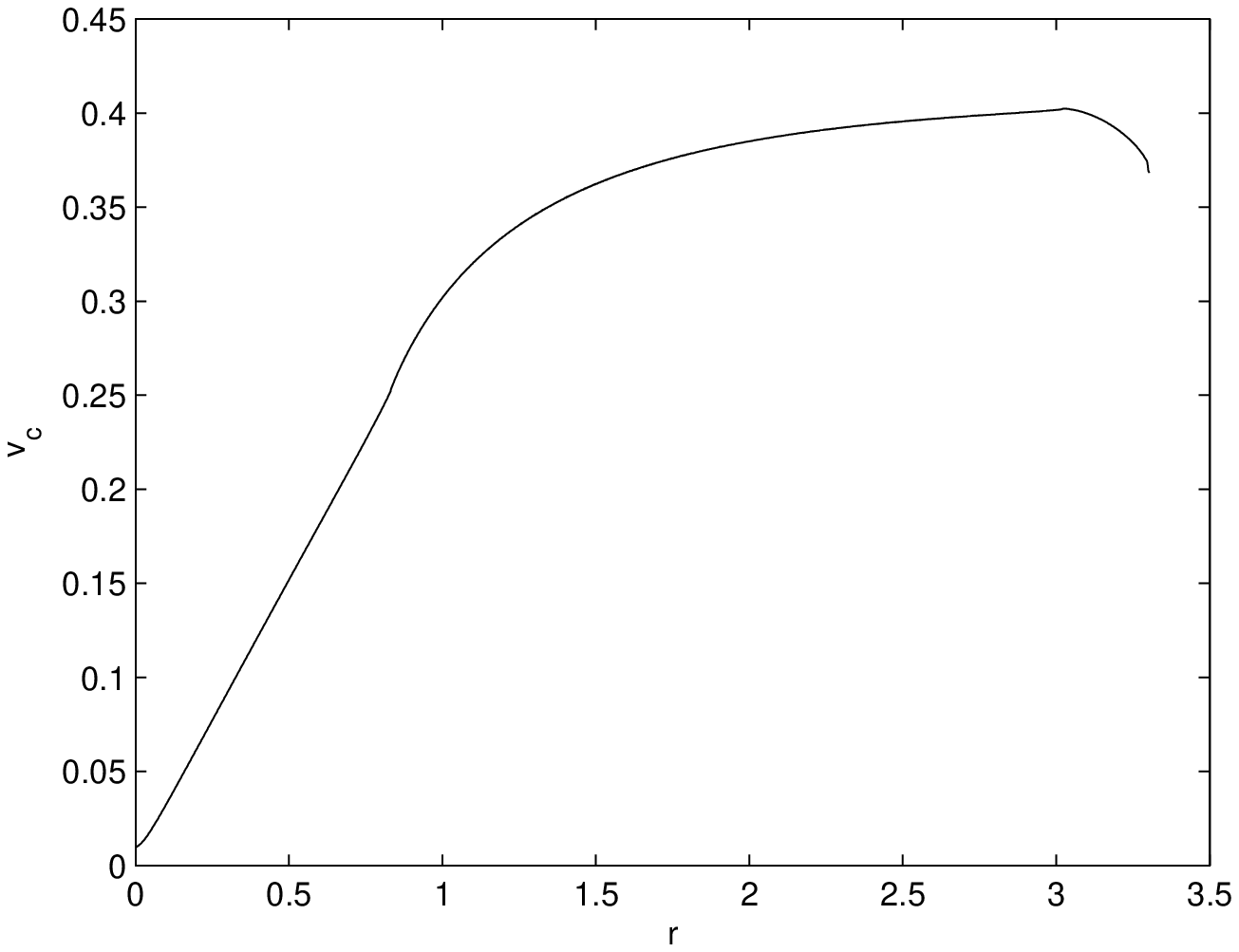}
  \caption{Rotation velocity $v_c$ versus radius $r$}
  \label{fig6:sub1}
\end{subfigure}%
\begin{subfigure}{.5\textwidth}
  \centering
  \includegraphics[width=1.0\linewidth]{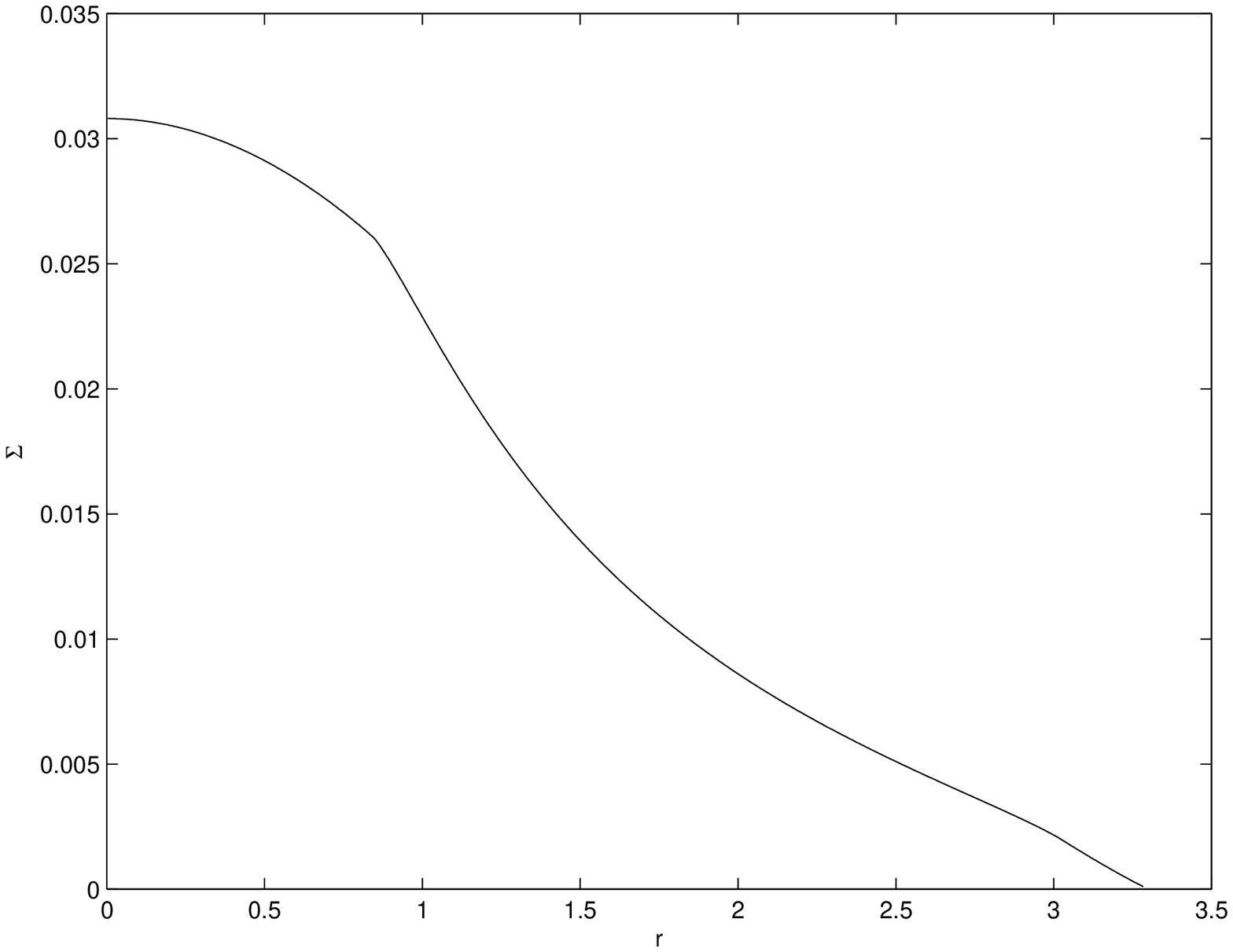}
  \caption{Density $\Sigma$ versus radius $r$}
  \label{fig6:sub2}
\end{subfigure}
\caption{$M=0.3,\;E_0=-0.1,\; l=0.0,\; Q=2.0$}
\label{fig6}
\end{figure}

\begin{figure}
\centering
\begin{subfigure}{.5\textwidth}
  \centering
  \includegraphics[width=1.0\linewidth]{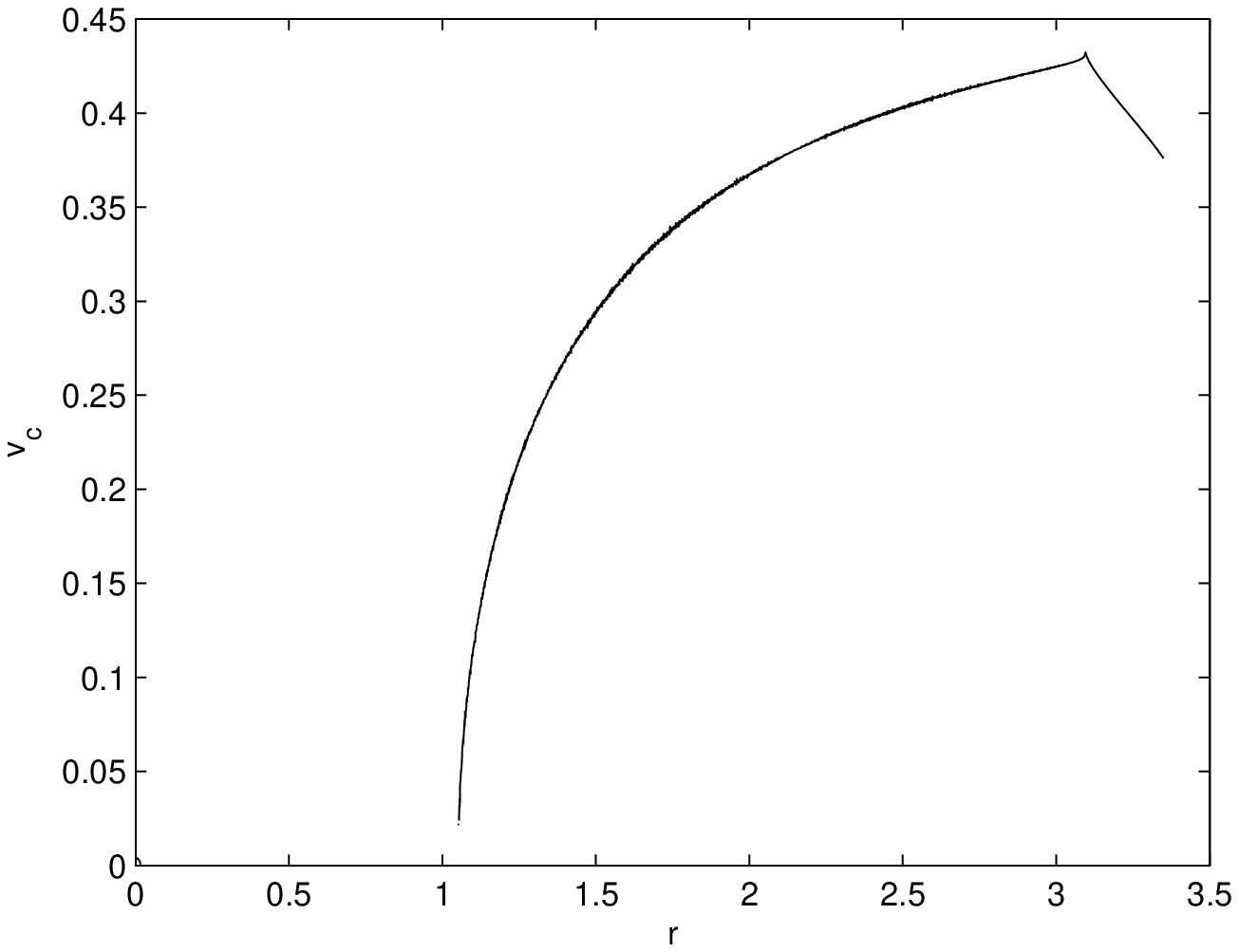}
  \caption{Rotation velocity $v_c$ versus radius $r$}
  \label{fig7:sub1}
\end{subfigure}%
\begin{subfigure}{.5\textwidth}
  \centering
  \includegraphics[width=1.0\linewidth]{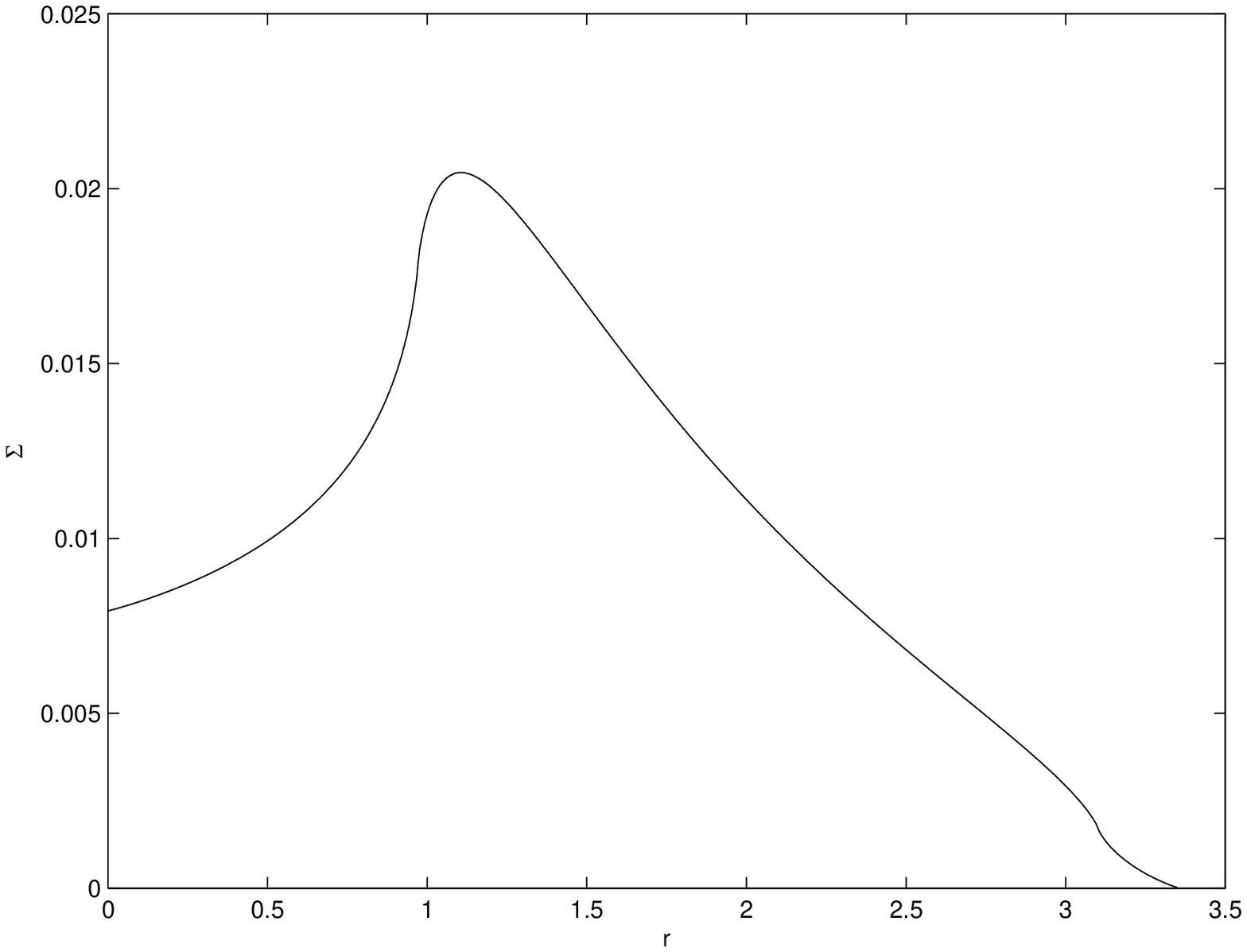}
  \caption{Density $\Sigma$ versus radius $r$}
  \label{fig7:sub2}
\end{subfigure}
\caption{$M=0.3,\;E_0=-0.1,\; l=-0.75,\; Q=2.0$}
\label{fig7}
\end{figure}

The influence of varying the parameter $l$ is studied in
Figures~\ref{fig1}, \ref{fig6}, and \ref{fig7}, where $l=1,0,-0.75$
respectively.

\begin{figure}[htbp]
\begin{center}
\scalebox{.50}{\includegraphics{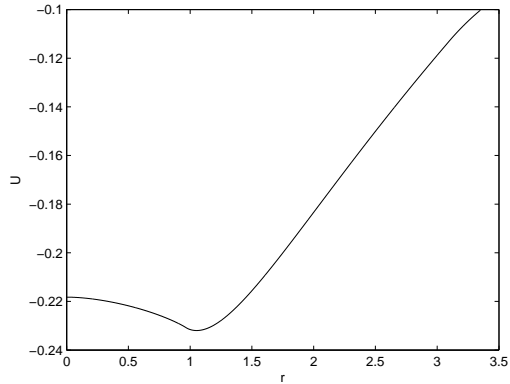}}
\end{center}
\caption{$U$ versus radius $r$ for the steady state in Figure~\ref{fig7}}
\label{fig8}
\end{figure}

\begin{figure}
\centering
\begin{subfigure}{.5\textwidth}
  \centering
  \includegraphics[width=1.0\linewidth]{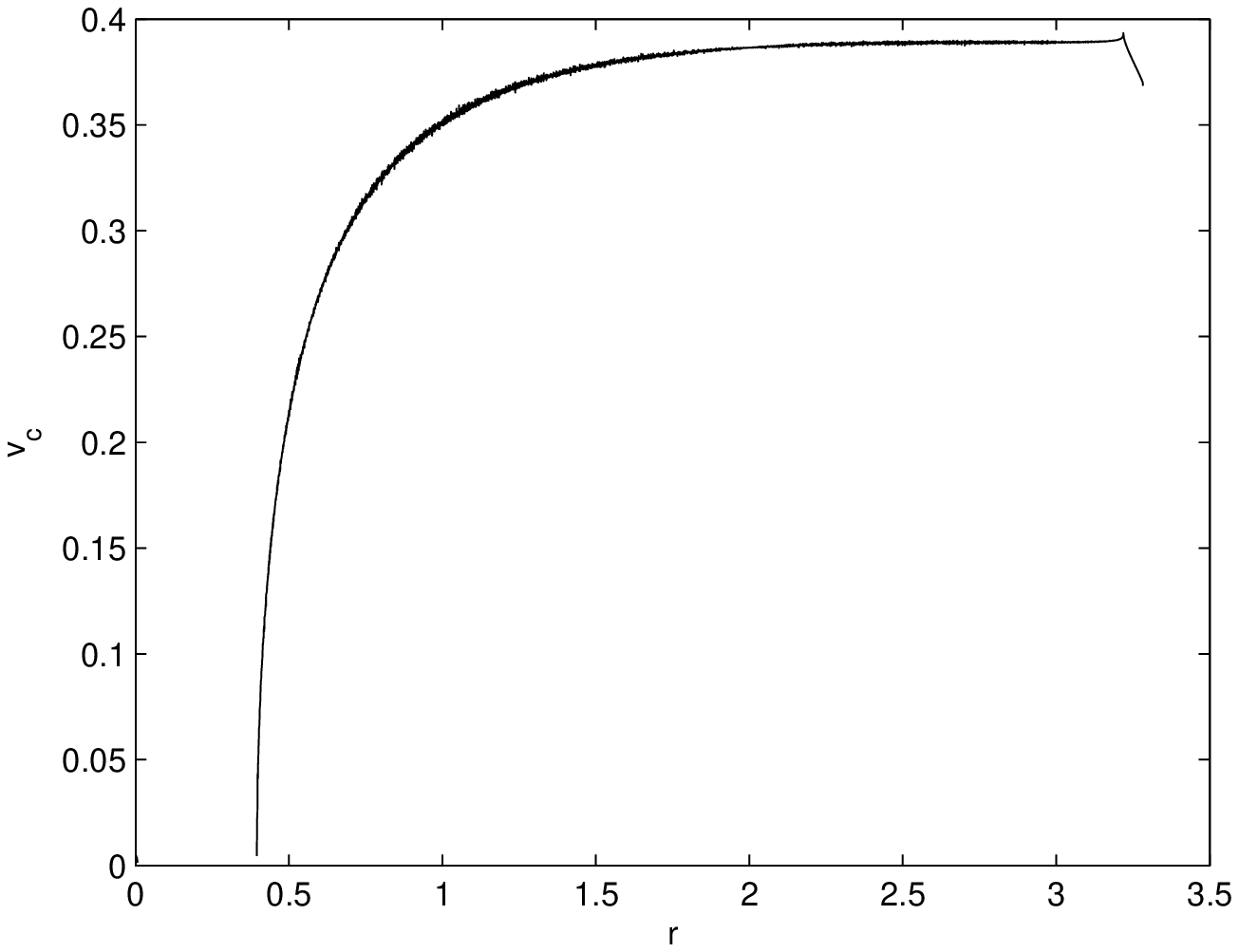}
  \caption{Rotation velocity $v_c$ versus radius $r$}
  \label{fig9:sub1}
\end{subfigure}%
\begin{subfigure}{.5\textwidth}
  \centering
  \includegraphics[width=1.0\linewidth]{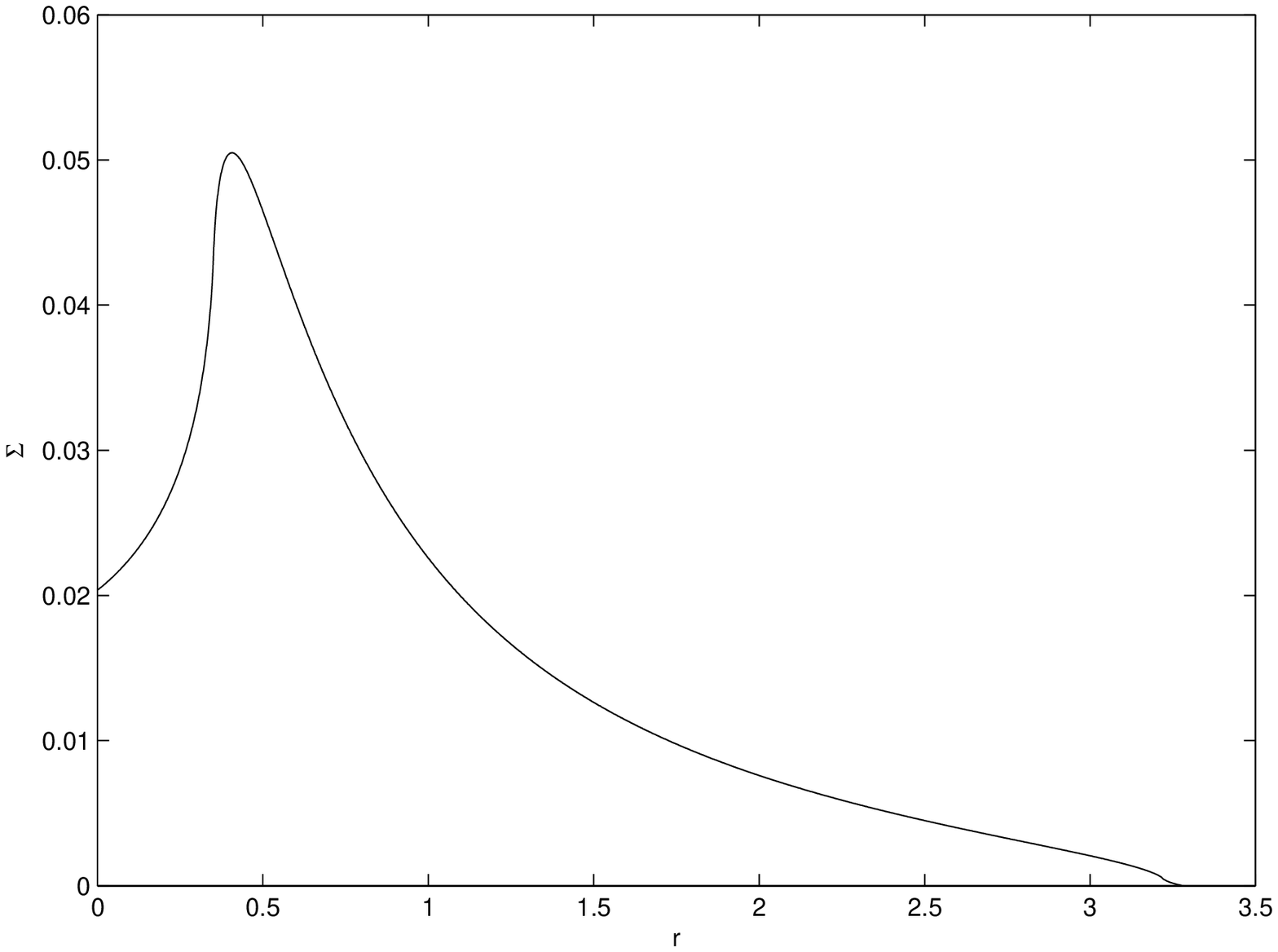}
  \caption{Density $\Sigma$ versus radius $r$}
  \label{fig9:sub2}
\end{subfigure}
\caption{$M=0.3,\;E_0=-0.1,\; l=-0.75,\; Q=4.1$}
\label{fig9}
\end{figure}

\begin{figure}[htbp]
\begin{center}
\scalebox{.50}{\includegraphics{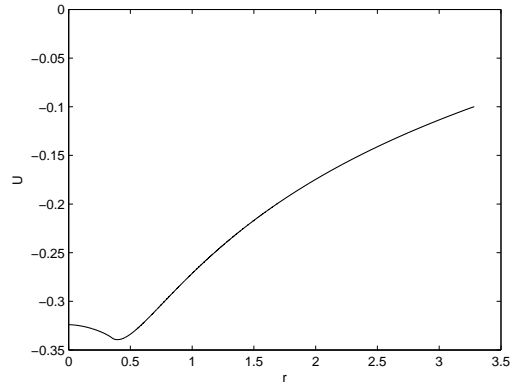}}
\end{center}
\caption{$U$ versus radius $r$ for the steady state in Figure~\ref{fig9}}
\label{fig10}
\end{figure}

We notice that when $l$ is decreased from $l=1$ to $l=0$ the regularity
of the graphs for $v_c$ and $\Sigma$ are affected and the slope of the
rotation curve increases. When $l$ is decreased further to $l=-0.75$
this effect is more pronounced. In particular neither the density $\Sigma$
in Figure~\ref{fig7} nor the potential $U$ in Figure~\ref{fig8} are
monotone.
In the domain where $U'(r)<0$, $v_c$ is not defined which is
clear from Figure~\ref{fig7:sub1}. Recall that in the spherically
symmetric case the potential is always increasing so the non-monotonicity
of $U$ is a particular feature of the axially symmetric Vlasov-Poisson system.
A similar example is given in Figures~\ref{fig9} and~\ref{fig10},
where $Q$ has been changed so that the rotation curve is approximately flat.

\begin{figure}[htbp]
\begin{center}
\scalebox{.50}{\includegraphics{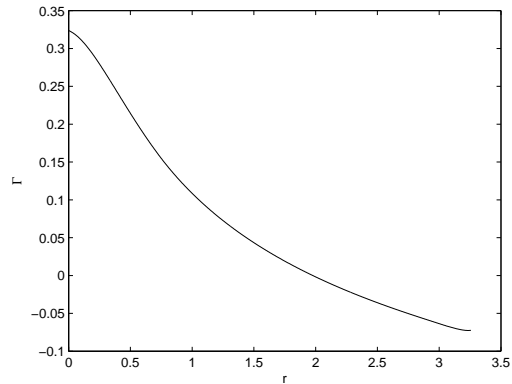}}
\end{center}
\caption{$\Gamma$ versus radius $r$ for the case in Figure~\ref{fig1}}
\label{fig11}
\end{figure}

\begin{figure}[H]
\begin{center}
\scalebox{.50}{\includegraphics{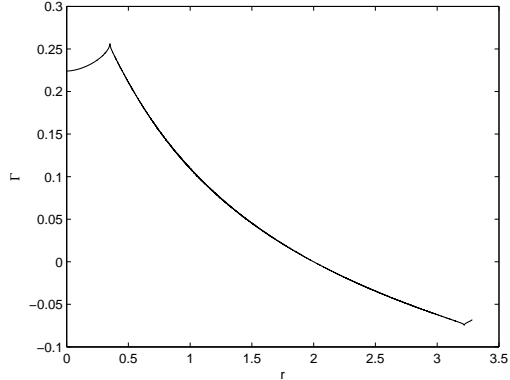}}
\end{center}
\caption{$\Gamma$ versus radius $r$ for the case in Figure~\ref{fig9}}
\label{fig12}
\end{figure}

In Theorem~\ref{nocircorbits} it was shown that there are no stars
in circular orbits in the neighborhood of the boundary of the steady state,
cf.~however the remark following the theorem.
Numerically it is straightforward to compute the range where stars
in circular orbits do not exist. By following the proof of
Theorem~\ref{nocircorbits} it is found that particles in circular
orbits do not exist if
\[
\Gamma:=E_0-\frac12 r\,U'(r)-U(r)<0.
\]
For the steady state given in Figure~\ref{fig1} $\Gamma$ is depicted in
Figure~\ref{fig11}.
In this case $R_b=3.26$ and $\Gamma$ is negative for $r>1.98$.
On circular orbits in the range $[1.98,3.36]$, i.e., in the range
$[0.6R_b,R_b]$, there are no particles
in the particle distribution given by $f$.
Hence, in the outer 40\% of the galaxy there are no
stars on circular orbits. Similarly, Figure~\ref{fig12} shows $\Gamma$
for the steady state in Figure~\ref{fig9}.
In this case $R_b=3.28$ and
we find that there are no stars on circular orbits
in the range $[2.0,3.28]$, i.e.,
in $[0.61R_b,R_b]$. Since we find in Figure~\ref{fig9} that $U'<0$ for
$r\leq 0.4$, there are of course no circular orbits at all when $r\leq 0.4$,
i.e., for $r\in [0,0.12R_b]$.

\begin{figure}[H]
\centering
\begin{subfigure}{.5\textwidth}
  \centering
  \includegraphics[width=1.0\linewidth]{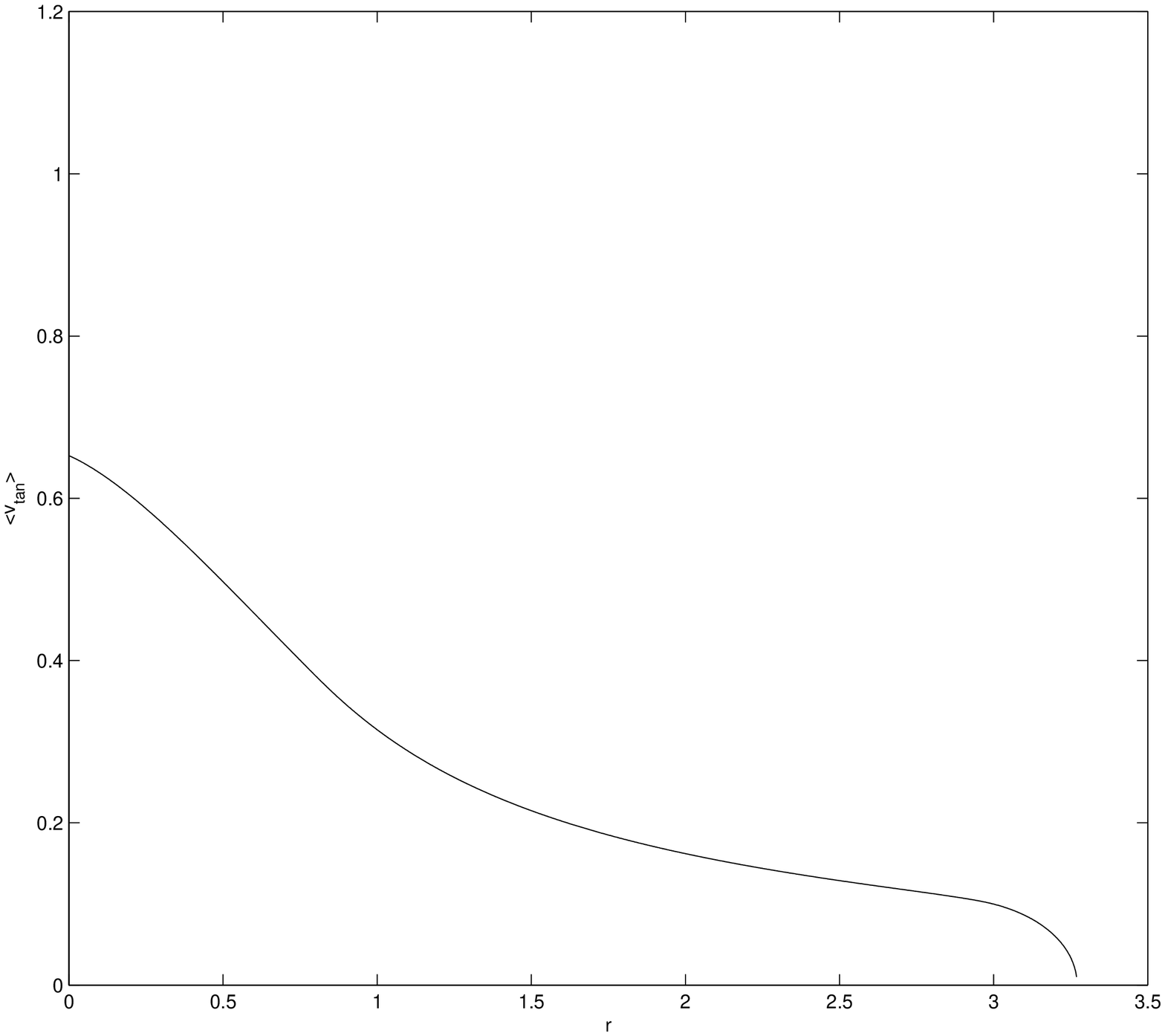}
  \caption{$\langle v_\mathrm{tan}\rangle$ versus radius $r$}
  \label{fig13n:sub1}
\end{subfigure}%
\begin{subfigure}{.5\textwidth}
  \centering
  \includegraphics[width=1.0\linewidth]{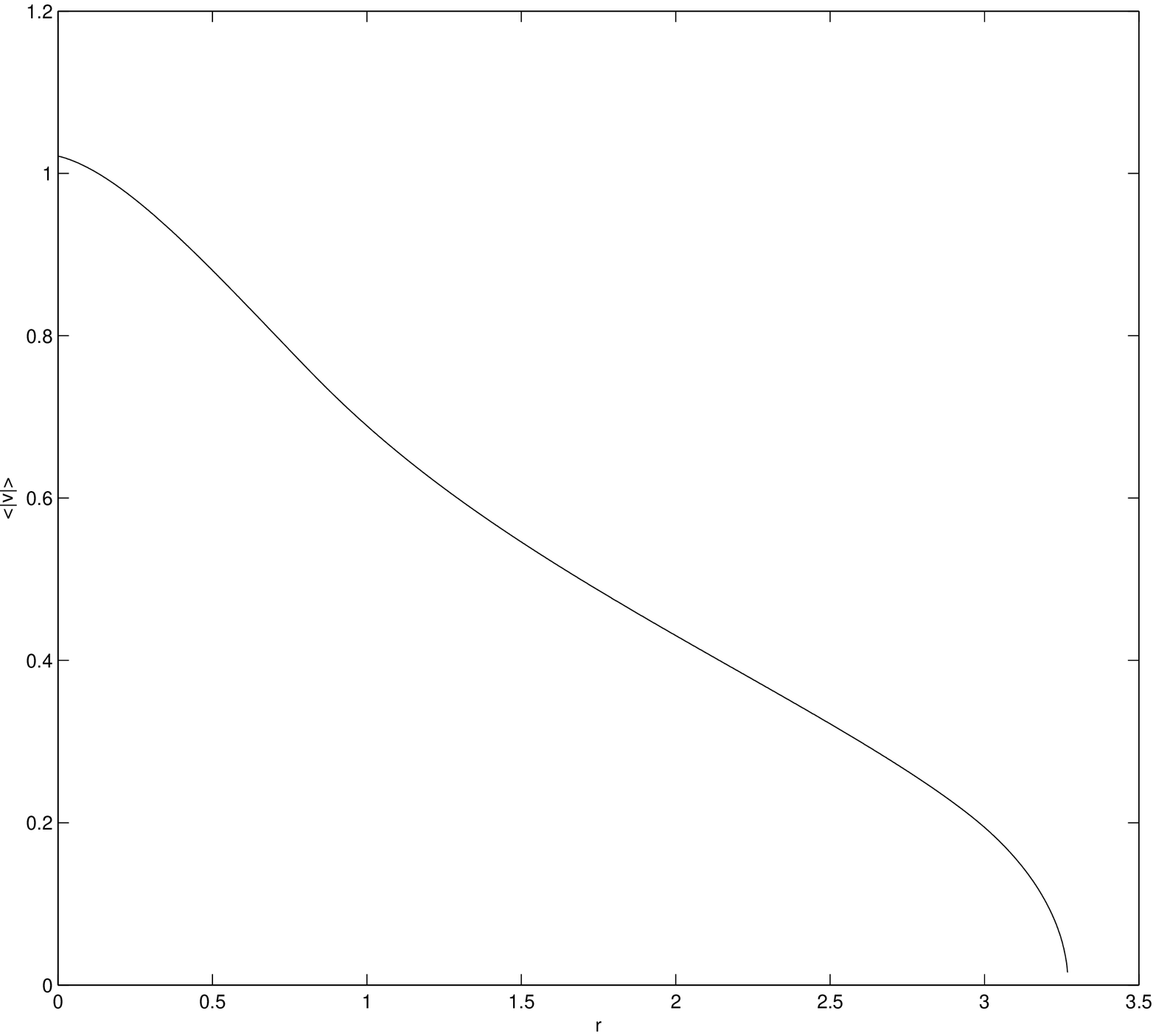}
  \caption{$\langle |v|\rangle $ versus radius $r$}
  \label{fig13n:sub2}
\end{subfigure}
\caption{$M=0.3,\;E_0=-0.1,\; l=1.0,\; Q=2.0$}
\label{fig13n}
\end{figure}

For the figures above we used Eqn.~\eqref{circorbit} to define the circular
velocity $v_c(r)$ at radius $r$. In the context of our model it is natural
to compute the averaged tangential velocity of the stars
from the phase space density $f$ instead, i.e.,
\[
\langle v_\mathrm{tan}(r)\rangle = \frac{1}{\Sigma(r)}\int_{\R^2} v_\mathrm{tan} f\, dv ,
\]
where $v_\mathrm{tan} = (x_1 v_2 - x_2 v_1)/r = L_z/r$
is the tangential component of the velocity
of a particle at $x$ with velocity $v$. It turns out that for the steady states
we constructed this quantity behaves
quite differently from $v_c$. As an example we plot
$\langle v_\mathrm{tan}(r) \rangle$ for the steady state shown in Figure~\ref{fig1},
cf.~Figure~\ref{fig13n:sub1}.
We also plot the average of $|v|$, and we observe that
both quantities behave quite differently from $v_c$.
This is in line with Theorem~\ref{nocircorbits}.
As noted in the remark after this theorem the observed
rotation curves often seem to be derived from the motion of
the interstellar gas, in particular this is the case for the data
which we try to fit in Section~\ref{compobs}.
Two possibilities for resolving the discrepancy between
the behaviour of $\langle v_\mathrm{tan}\rangle$ and of $v_c$
in our models come to mind. One possibility is to add a second
particle species to the model which is to represent the gas
and from which the averaged tangential velocity $\langle v_\mathrm{tan} \rangle$
would then be computed. Another possibility is to search for ansatz functions
for which the difference between $\langle v_\mathrm{tan}\rangle$ and $v_c$
is small. Both possibilities are currently under investigation.
It is of interest to note from Figure~\ref{fig13n}
that in the central part of the particle distribution we construct there
are stars which move fast compared to the rotation velocity $v_c$
and which in particular move on excentric orbits. In view of the stability
of flat galaxies this may be a positive feature of the model,
cf.~(Kalnajs 1976).

An interesting question is how much mass is needed to obtain the circular rotation
velocities in the outer region of the steady state if a Keplerian approach is used.
Consider for instance the case depicted in
Figure~\ref{fig1}. At $r=3.0$ we have from the numerical simulation that
$m(3.0)=0.295$, and $v_c=0.39$. The mass $M_{K}$ required to obtain
$v_c=0.39$ using the Kepler formula
\[
v_c^2=\frac{M_{K}}{r},
\]
is thus, in this case, $M_{K}=r v_c^2=0.46$. Hence,
\[
\frac{M_{K}}{m(3.0)}=\frac{0.46}{0.295}=1.56,
\]
which implies that more than 50\% additional mass is required to explain
the rotation curve obtained in Figure~\ref{fig1}. Analogously, for the
steady state in Figure~\ref{fig7}, $v_c=0.43$ at $r=3.0$, and $m(3.0)=0.29$,
which implies that $M_{K}=0.56$ and
\[
\frac{M_{K}}{m(3.0)}=\frac{0.56}{0.29}=1.93,
\]
so that more than 90\% additional mass is required using the Kepler formula.

\section{Comparison to observations}\label{compobs}
\setcounter{equation}{0}
In this section we consider data for some of the spiral galaxies
studied by Verheijen \& Sancisi (2001)
which belong to the Ursa Major cluster.
The aim is to find solutions of the flat Vlasov-Poisson
system which match these data.

In the introduction we stated the Vlasov-Poisson system
in dimensionless form, but in order to compare our results with
observations we need to attach proper units.
Since the gravitational constant $G$ is the only physical
constant which enters the system and since we normalized this to unity in
\eqref{poisson} we can choose any set of units for time, length, and mass
in which the numerical value for $G$ equals unity.
When we then try to fit the numerical predictions to
the observations of a certain galaxy we first choose the
unit for length such that the numerically obtained value for
its radius corresponds to the observed radius. Then we choose the unit
of time (and hence velocity) in such a way that the numerically predicted
rotation curve fits the observations---of course this only works
if the mathematical model we consider produces a rotation curve
with the proper shape, which we try to achieve by varying
the parameters in the ansatz \eqref{ansatz1}.
Once the units for length and time are chosen in this way, the condition
$G=1$ fixes the unit for mass, and we can for example
transform the numerically obtained value for the mass into
a predicted mass of the galaxy under consideration in units of
solar masses.

In the pictures below we have normalized the radius so that the boundary occurs at
$r=1$. However, we choose not to identify the radius of the last observation, i.e.,
the largest radius of the observational data, with the boundary of the support $R_b$
of a solution since the density vanishes at the boundary. Instead we identify it with
a radius clearly within the support and we choose to identify it with $\lambda R_b$,
where $\lambda\approx 0.96$.
\begin{figure}[H]
\begin{center}
\scalebox{.45}{\includegraphics{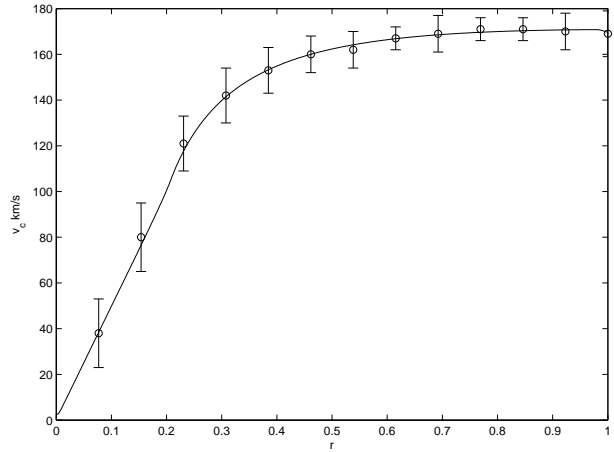}}
\end{center}
\caption{The galaxy NGC3877. Observational data (circles) and VP solution (solid line).}
\label{fig13}
\end{figure}
\begin{figure}[H]
\begin{center}
\scalebox{.45}{\includegraphics{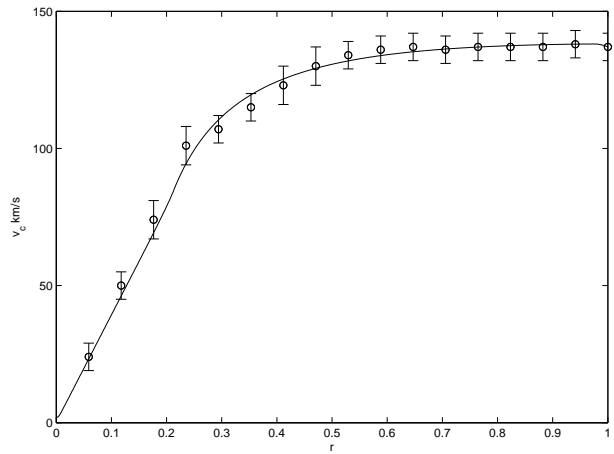}}
\end{center}
\caption{The galaxy NGC3917. Observational data (circles) and VP solution (solid line).}
\label{fig14}
\end{figure}
The measured rotation curves for the galaxies NGC3877, NGC3917, NGC3949, and NGC4010 are
depicted by open circles in Figures~\ref{fig13}, \ref{fig14},  \ref{fig15},
and \ref{fig16} respectively. The uncertainties in the observational data are tabulated
in (Verheijen \& Sancisi~2001) and are shown as error bars.
\begin{figure}[H]
\begin{center}
\scalebox{.45}{\includegraphics{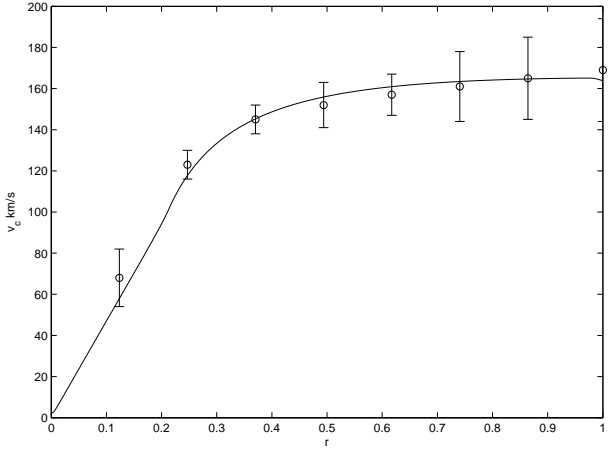}}
\end{center}
\caption{The galaxy NGC3949. Observational data (circles) and VP solution (solid line).}
\label{fig15}
\end{figure}
\begin{figure}[H]
\begin{center}
\scalebox{.45}{\includegraphics{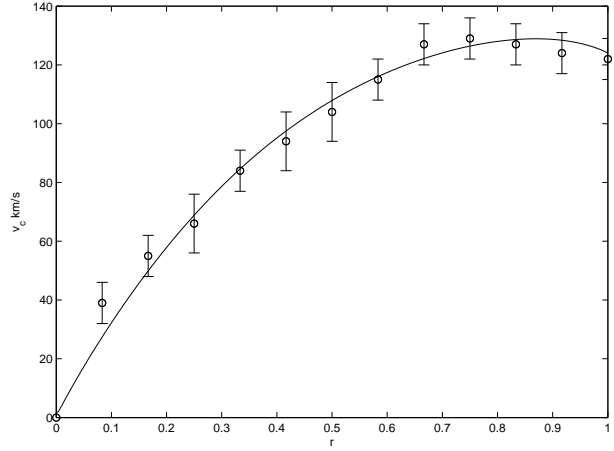}}
\end{center}
\caption{The galaxy NGC4010. Observational data (circles) and VP solution (solid line).}
\label{fig16}
\end{figure}
The solid curve in each of these figures is the rotation curve given by the solution
of the flat Vlasov-Poisson system using the ansatz~(\ref{ansatz1}). The solution is
then scaled by the procedure described above. The parameter values in the
ansatz~(\ref{ansatz1}) are as follows.
In all the four cases $k=0, M=0.3$ and $E_0=-0.1$.
In Figure~\ref{fig13} $l=0$ and $Q=2.40$,
in Figure~\ref{fig14} $l=0$ and $Q=2.30$,
in Figure~\ref{fig15} $l=0$ and $Q=2.35$,
and in Figure~\ref{fig16} $l=1$ and $Q=0.65$.
Taking into account the uncertainties of the observational data
we conclude that the solutions agree very well with the observations.

As explained above we can predict the total mass of these galaxies
from the numerical results, and the corresponding predictions are
tabulated in Table~1.
\begin{table}[H]
\begin{centering}
\begin{tabular}{|l|r|}
\hline
galaxy & predicted mass\\
       & in solar masses\\
\hline
NGC3877 & $4.7 \cdot 10^{10} M_{\odot}$\\
NGC3917 & $4.0 \cdot 10^{10} M_{\odot}$\\
NGC3949 & $3.5 \cdot 10^{10} M_{\odot}$\\
NGC4010 & $2.2 \cdot 10^{10} M_{\odot}$\\
\hline
\end{tabular}
\caption{Predicted masses}
\end{centering}
\end{table}
The masses we obtain agree well with the ones obtained
by Gonz\'{a}lez, Plata \& Ramos-Caro~(2010) with a completely
different fitting approach.

\textbf{Acknowledgements. }
The authors sincerely thank an anonymous referee for several very helpful
and constructive suggestions. They also thank  Roman Firt for teaching them
how to use the gsl package for computing the elliptic integral in
equation~(\ref{UK}),
and Markus Kunze for pointing out the
reference (Dejonghe~1986). The material presented here
is based on work supported by the
National Science Foundation under Grant No. 0932078~000, while the
first author was in residence at the Mathematical Sciences Research
Institute in Berkeley, California, during the fall of 2013.
The first author wants to express his gratitude for the invitation to MSRI.


\bigskip

\noindent
{\bf \Large References}



\smallskip \noindent
Amorisco N.~C., Bertin G., 2010,
AIP, 1242, 288

\smallskip \noindent
Batt J., Faltenbacher W., Horst E., 1986,
Arch.\ Rat.\ Mech.\ Anal., 93, 159

\smallskip \noindent
Binney J., Tremaine S., 1987,
Galactic Dynamics, Princeton Univ.\ Press, Princeton, NJ

\smallskip \noindent
Blok W.~J.~G~de, Walter F., Brinks E., Trachternach C., Oh S.-H.,
Kennicutt R.~C.\ Jr., 2008,
Astron.\ J., 136, 2648

\smallskip \noindent
Bosma A., 1981,
Astron.\ J., 86, 1825

\smallskip \noindent
Chavanis P.~H., 2003,
A\&A, 401, 15

\smallskip \noindent
Dejonghe H., 1986,
Physics reports, 133, 217

\smallskip \noindent
Famaey B., McGaugh S.~S., 2012,
Living Rev.\ Relativity, 15

\smallskip \noindent
Firt R., Rein G., 2006,
Analysis, 26, 507

\smallskip \noindent
Fuchs B., 2001,
Dark Matter in Astro- and Particle Physics.
Proc.\ Int.\ Conf.\ DARK 2000, Heidelberg, p.~25

\smallskip \noindent
Gentile G., Salucci P., Klein U., Vergani D., Kalberla P., 2004,
MNRAS, 351, 903

\smallskip \noindent
Gonz\'{a}lez G.~A., Plata S.~M.~, Ramos-Caro J., 2010,
MNRAS, 404, 468

\smallskip \noindent
Hur\'{e} J.-M., 2005,
A\& A, 434, 1

\smallskip \noindent
Kalnajs A.~J., 1972,
Astrophys.\ J., 175, 63

\smallskip \noindent
Kalnajs A.~J., 1976,
Astrophys.\ J., 205, 751

\smallskip \noindent
Lynden-Bell D., 1967,
MNRAS, 136, 101

\smallskip \noindent
Moffat J.~W., 2006,
J.\ of Cosmology and Astroparticle Physics, 03, 004

\smallskip \noindent
Nguyen P.~H., Lingam M.,2013,
MNRAS, 436, 2014

\smallskip \noindent
Pedraza J.~F., Ramos-Caro J., Gonz\'{a}lez G.~A.,2008,
MNRAS, 390, 1587

\smallskip \noindent
Ramming T., Rein G., 2013,
SIAM J.\ Math.\ Anal., 45, 900

\smallskip \noindent
Ramos-Caro J., Agn C.~A., Pedraza J.~F., 2012,
Phys.\ Rev.\ D, 86, 043008

\smallskip \noindent
Rein G., 2007,
in Dafermos C.~M., Feireisl E., eds,
Handbook of Differential Equations,
Evolutionary Equations, 3

\smallskip \noindent
Rein G., 1999,
Commun.\ Math.\ Phys., 205, 229

\smallskip \noindent
Rein G., 2013,
preprint (arXiv:1312.3765)

\smallskip \noindent
Rein G., Rendall A.~D., 2000,
Math.\ Proc.\ Camb.\ Phil.\ Soc., 128, 363

\smallskip \noindent
Rodrigues D.~C., Letelier P.~S., Shapiro I.~L., 2010,
J.\ of Cosmology and Astroparticle Physics, 04, 020

\smallskip \noindent
Roos M., 2010,
preprint (arXiv:1001.0316)

\smallskip \noindent
Rubin V.~C., Ford J., Thonnard W.~K., Burnstein D., 1982,
Astron.\ Rep., 261, 439

\smallskip \noindent
Salucci P., 2007,
Proc.\ IAU, 244, 53

\smallskip \noindent
Schulz E., 2012,
Astrophys.\ J., 747, 106

\smallskip \noindent
Verheijen M.~A. Sancisi R., 2001,
A\& A, 370, 765

\end{document}